\documentclass[twocolumn,english,pra,showpacs,superscriptaddress,longbibliography]{revtex4-1}
\usepackage{amsfonts}
\usepackage{amsmath}
\usepackage{graphics}
\usepackage{amssymb}
\usepackage{physics}
\usepackage[colorlinks, citecolor=blue,linkcolor=red]{hyperref}
\usepackage{color}
\usepackage{hyperref}
\usepackage{textcomp}
\usepackage[rightcaption]{sidecap}
\usepackage{subfigure}
\usepackage{float,csquotes,mathtools}
\usepackage[rightcaption]{sidecap}
\usepackage{graphicx}
\usepackage{lipsum}

\begin{document}
\title{Goos–H\"{a}nchen Shift in $\mathcal{PT}$-Symmetric and Passive Cavity Optomechanical Systems}
\author{Shah Fahad}
\affiliation{Department of Physics, Zhejiang Normal University, Jinhua, Zhejiang 321004, China}
\author{Gao Xianlong}
\email {gaoxl@zjnu.edu.cn}\affiliation{Department of Physics, Zhejiang Normal University, Jinhua, Zhejiang 321004, China}

\setlength{\parskip}{0pt}
\setlength{\belowcaptionskip}{-10pt}
\begin{abstract}
We theoretically investigate the control of the Goos–H\"{a}nchen shift (GHS) of a reflected weak probe field in both parity–time ($\mathcal{PT}$)-symmetric and conventional optomechanical systems. The proposed scheme consists of a single optomechanical platform where a passive optical cavity is coupled to an active mechanical resonator, in contrast to standard passive–passive configurations. Analysis of the eigenfrequency spectrum reveals the emergence of an exceptional point under balanced gain–loss conditions at a tunable effective optomechanical coupling strength. Using the transfer‑matrix method combined with stationary‑phase analysis, we examine the GHS across broken and unbroken $\mathcal{PT}$ phases and compare it with that in the conventional system. The lateral shift exhibits strong phase dependence: it is markedly enhanced in the unbroken regime relative to both the broken phase and the passive configuration. We further show that the GHS can be actively tuned through the cavity detuning and the intracavity medium length. These results provide a controlled means for manipulating beam shifts in optomechanical systems and suggest pathways toward tunable photonic components and precision optical sensing.
\end{abstract}
\date{\today}
\maketitle
\section{Introduction}
Propelled by recent technological advances, cavity optomechanics—the study of interactions between light and mechanical oscillators via radiation-pressure forces—has emerged as a leading platform for both fundamental and applied research~\cite{Aspelmeyer2014, Aspelmeyer-book-2014, Kippenberg}. Recent experiments have demonstrated a range of quantum phenomena in cavity optomechanics—from ponderomotive squeezing~\cite{PurdyPRX, Brooks2012} and ground-state cooling~\cite{Chan2011} to quantum nondemolition measurements~\cite{PontinPRA}. Leveraging these advances, phenomena such as optomechanically induced amplification (OMIAMP)~\cite{Massel2011, Zhou2013, Monifi2016}, absorption (OMIA)~\cite{Massel2012}, and transparency (OMIT)~\cite{Stefan2010, Karuza2013} have been realized, enabling precise manipulation of light in micro- and nano-fabricated structures at room temperature.

Parity-time ($\mathcal{PT}$)-symmetric Hamiltonians—despite being non-Hermitian—can exhibit entirely real spectra~\cite{BenderPRL}. Their experimental demonstration has attracted significant attention~\cite{Chtchelkatchev-PRL, Bittner2012, Bender2013, Shi2016}, particularly in optical systems with balanced loss and gain~\cite{Ruschhaupt2005, Bender2004, Bender2013TWO, El2007theory}. These systems exhibit a phase transition from the broken $\mathcal{PT}$ phase (complex spectrum) to the unbroken $\mathcal{PT}$ phase (real spectrum) at the exceptional point ($\mathrm{EP}$), where eigenvalues and eigenvectors coalesce~\cite{Bender2013, Iorsh-PRL, Liu_2017}. Based on these advances, $\mathcal{PT}$-symmetry phase transitions have also been experimentally observed in microcavities~\cite{Peng2014, Peng2014Science}, waveguides~\cite{Ruter2010, KONOTOP2012,Feng2013, LinPRL}, and active LRC circuits~\cite{Schindle2011}. Moreover, the role of $\mathcal{PT}$ symmetry and its breaking has been extensively investigated in coupled-cavity optomechanics. In these systems, a $\mathcal{PT}$-symmetric configuration is typically realized by coupling an active (gain) optical cavity without a mechanical mode to a passive (lossy) cavity that supports one~\cite{Jiao-2016, Li2016, Liu2016PRL, He-2016}. This phase transition has been shown to enable a range of theoretically predicted phenomena, including ultralow-threshold optical chaos~\cite{Lu-2015}, $\mathcal{PT}$-induced amplification~\cite{He-2019}, inverted OMIT~\cite{Jing2015}, coherent absorption~\cite{Zhang-2017}, and phonon lasing~\cite{Jing-PRL, Liu_2017, Zhang2018, Zhong2019}. By leveraging coherent phonon manipulation~\cite{OConnell2010, Fan2015, Okamoto2013, Mahboob2012} with an additional optical control field, a tunable optomechanical coupling can be established between the lossy optical cavity and the gain-enhanced mechanical resonator, effectively realizing a $\mathcal{PT}$-symmetric–like optomechanical system~\cite{Liu_2017, Xu-2021}. Based on this platform, we theoretically analyze the Goos–H\"{a}nchen shift (GHS) as a distinct and insightful signature of the system’s optical response in both the broken and unbroken $\mathcal{PT}$-symmetric phases, as well as in a conventional system.

GHS~\cite{Goos-1943, Goos-1947} is the lateral displacement of a light beam relative to its geometric center that occurs upon total internal reflection at the interface between two media.
This phenomenon underscores the superior accuracy of wave optics over ray optics in describing light propagation within dielectric media~\cite{Renard1964, Bliokh2013, Schomerus2006}, and enables significant applications in interferometry and optical sensing~\cite{deFornel2001, Yin2006, Soboleva-2012}. Owing to its origin as an interference effect, the GHS is not limited to optical systems but also manifests in a wide range of wave phenomena, including neutrons~\cite{Hann-PRL}, electron waves~\cite{Beenakker-PRL, Wu-PRL}, spin waves~\cite{Dadoenkova-2012}, Weyl media~\cite{Jiang-PRL, Chattopadhyay2019, Liu-2020}, and matter waves~\cite{Huang-PRL, Lee-2014}. Moreover, the GHS has been explored in diverse $\mathcal{PT}$-symmetric systems, such as atomic vapors~\cite{Han2021}, layered structures~\cite{Yue2021, Ma2017}, photonic crystals~\cite{Zhang2022, Longhi2011}, atomic ensembles in optical cavities~\cite{Ziauddin2015}, and cavity magnomechanical systems~\cite{fahad2025}.

Cavity optomechanical systems provide a compelling platform for tailoring the GHS. Through radiation-pressure interaction between an optical and a mechanical mode, these systems enable the generation of sharp, tunable phase-dispersion and interference features—such as transparency and absorption windows—that critically determine both the magnitude and the sign of the GHS for a reflected probe field~\cite{Muhib-2019, Ghaisuddin_2021, Anwar-2020}. Despite these advances, the GHS has not yet been systematically explored or compared between the broken and unbroken $\mathcal{PT}$-symmetric phases in a $\mathcal{PT}$-symmetric optomechanical system comprising a passive optical cavity coupled to an active mechanical resonator, nor compared with a conventional system on the same physical platform. This gap is addressed in the present study.

We consider a $\mathcal{PT}$-symmetric optomechanical system that spans the broken $\mathcal{PT}$ phase, exceptional point ($\mathrm{EP}$), and the unbroken $\mathcal{PT}$ phase. The platform consists of an optical cavity with one fixed and one movable mirror, configured such that a passive optical cavity couples to an active mechanical resonator. This contrasts with the conventional optomechanical system, where both the optical cavity and mechanical resonator are passive. Under conditions of gain-loss balance, the system exhibits an $\mathrm{EP}$ at a critical value of the effective optomechanical coupling strength.

Using the transfer-matrix method, we first calculate the reflection coefficient and then apply the stationary-phase approach to systematically analyze the GHS of the reflected probe field across the different $\mathcal{PT}$-symmetric phases as well as the conventional system. Our analysis reveals a strong phase dependence: the GHS is significantly enhanced in the unbroken $\mathcal{PT}$ relative to both the broken phase and the conventional case. This finding aligns with prior observations of enhanced beam shifts in other $\mathcal{PT}$-symmetric systems~\cite {Ziauddin2015, fahad2025}. Furthermore, we demonstrate active control of the GHS across the $\mathcal{PT}$ phases and conventional configuration by tuning the probe-cavity deutning and intracavity length. These results establish a pathway for precise control of the GHS in optomechanical systems, introducing a versatile platform for tunable microwave photonic devices and high-sensitivity optical sensing.

The paper is organized as follows. Section II introduces the theoretical model: (A) starting from the system Hamiltonian, we derive the linearized Heisenberg-Langevin equations; (B) we discuss the effective Hamiltonian and $\mathcal{PT}$-symmetry; (C) we analytically compute the optical susceptibility; and (D) we apply the stationary phase method to evaluate the Goos–H\"{a}nchen shift. Section III presents numerical results and discussion, and Section IV concludes.
\begin{figure}
\centering
\includegraphics[width=0.95\linewidth]{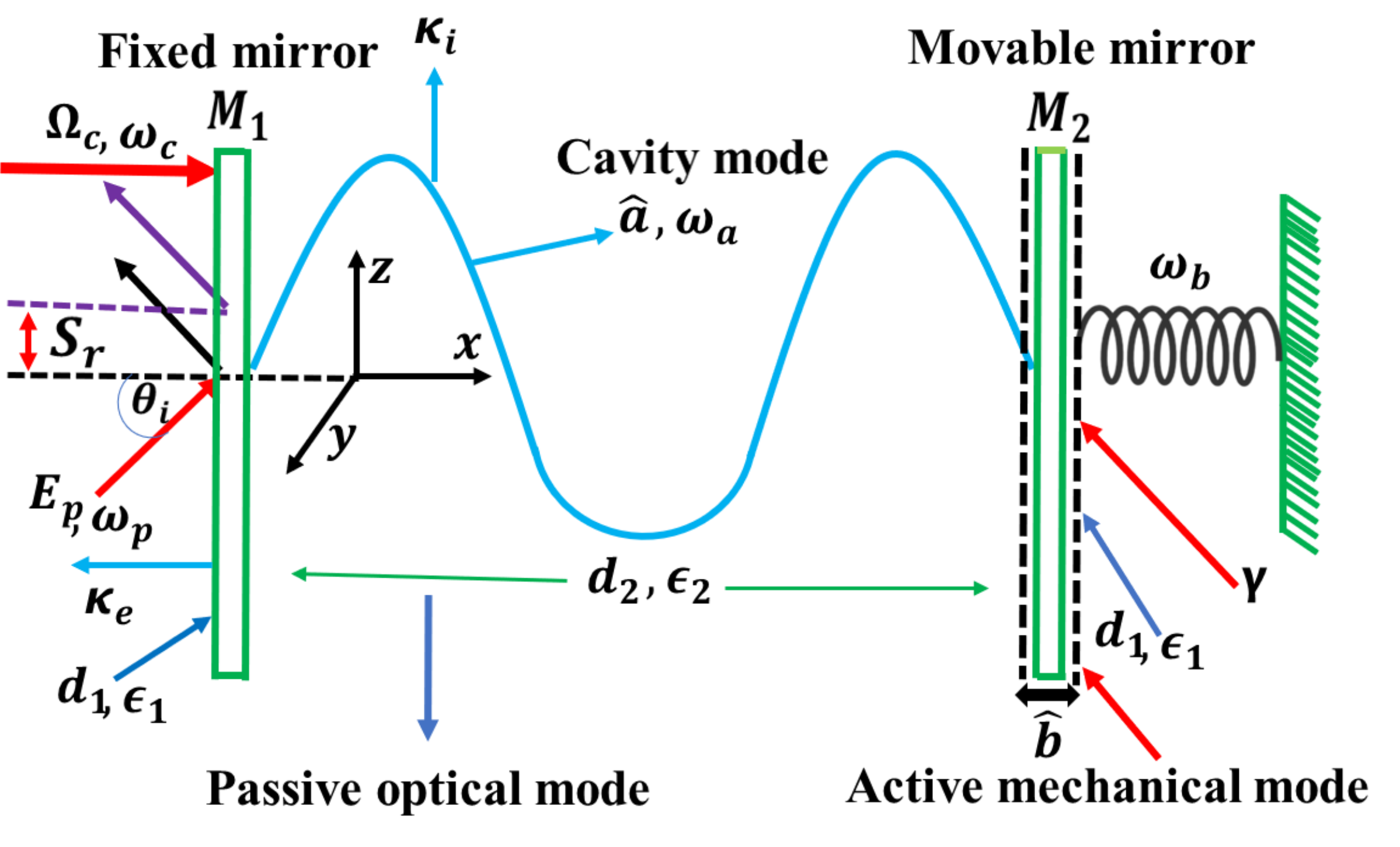}
\caption{Schematic illustration of a $\mathcal{PT}$-symmetric optomechanical system comprising a passive optical cavity ($\hat{a}$, resonance frequency $\omega_{a}$, total decay rate $\kappa = \kappa_i + \kappa_e$) and an active mechanical resonator ($\hat{b}$, resonance frequency $\omega_{b}$, gain rate $\gamma$). The cavity is formed by two nonmagnetic mirrors, $M_1$ (fixed) and $M_2$ (movable). A strong control field (amplitude $\Omega_{c}$, frequency $\omega_{c}$) drives the cavity, which exhibits intrinsic and external coupling decay rates $\kappa_i$ and $\kappa_e$, respectively. A transverse electric (TE) polarized probe field (amplitude $E_{p}$, frequency $\omega_{p}$) is incident on mirror $M_1$ at an angle $\theta_i$. The lateral displacement of the reflected probe field, known as the Goos–H\"{a}nchen shift, is denoted by $S_r$.}
\label{fig1}
\end{figure}
\section{System and Hamiltonian}
\subsection{$\mathcal{PT}$-symmetric and conventional optomechanical systems}
We consider a $\mathcal{PT}$-symmetric optomechanical system comprising a single-mode passive optical cavity ($\hat{a}$, resonance frequency $\omega_a$, total decay rate $\kappa$) coupled to an active mechanical resonator ($\hat{b}$, resonance frequency $\omega_b$, gain rate $\gamma$), as shown in Fig.~\ref{fig1}. In contrast, a conventional optomechanical setup couples a passive optical cavity to a passive mechanical resonator. The optical cavity mode is coherently driven by a strong control field (amplitude $\Omega_{c} = \sqrt{P_{c}\kappa_{e}/\hbar\omega_{c}}$, power $P_{c}$, coupling decay rate $\kappa_{e}$, frequency $\omega_c$) and a probe field (amplitude $E_{p} = \sqrt{P_{p}\kappa_{e}/\hbar\omega_{p}}$, power $P_p$, frequency $\omega_p$, incident angle $\theta_i$). The cavity is formed by two nonmagnetic mirrors: a fixed mirror $M_1$ and a movable mirror $M_2$, each with thickness $d_1$ and permittivity $\epsilon_1$. Mirror $M_1$ is partially reflective, while $M_2$ is perfectly reflective. The mirrors are separated by a distance $d_2$, enclosing an intracavity medium with effective permittivity $\epsilon_2$. 

In the frame rotating at frequency $\omega_{c}$, the system Hamiltonian is expressed as~\cite{Liu_2017}
\begin{equation}
\begin{split}
\hat{\mathcal{H}} &= \hbar\Delta_{a}\hat{a}^{\dagger}\hat{a} + \hbar\omega_{b}\hat{b}^{\dagger}\hat{b} - \hbar g_0 \hat{a}^{\dagger}\hat{a}(\hat{b}^{\dagger}+\hat{b})\\
&+ i\hbar(\Omega_{c}\hat{a}^{\dagger} + E_{p}\hat{a}^{\dagger}e^{-i\delta_{p} t} - \mathrm{H.c}.),\label{M-H}
\end{split}
\end{equation}
where $\Delta_{a} = \omega_{a} - \omega_{c}$ and $\delta_{p}=\omega_{p}-\omega_{c}$ are the cavity–control and probe–control detunings, respectively. Here, $\hat{a}$ ($\hat{a}^\dagger$) and $\hat{b}$ ($\hat{b}^\dagger$) are the annihilation (creation) operators of the cavity and mechanical modes, respectively, and $g_{0}$ corresponds to the single-photon optomechanical coupling strength. The coupling between the external field and the cavity mode is quantified by the parameter $\eta = \kappa_e/(\kappa_i + \kappa_e)$, where $\kappa_i$ and $\kappa_e$ are the intrinsic and external cavity decay rates, and $\kappa=\kappa_i+ \kappa_e$ is the total cavity decay rate. These rates can be continuously tuned in experiments. The system operates in the undercoupled regime for $\eta \ll 1$, in the overcoupled regime for $\eta \simeq 1$, and reaches critical coupling when $\eta = 1/2$~\cite{Cai-PRL, Spillane-PRL}.

The semiclassical Heisenberg-Langevin equation (HLE) for the operator $\hat{O} \in \{\hat{a},\hat{b}\}$ takes the form:
\begin{equation}
\frac{d\hat{O}}{dt} =  \frac{i}{\hbar} {[\hat{\mathcal{H}}, \hat{O}]} -\frac{\zeta}{2}\hat{O}+ \mathcal{N},\label{Generic form}  
\end{equation}
where $\zeta$ represents the decay ($\zeta>0$) or gain rate ($\zeta<0$), $[\hat{O}, \hat{O}^\dagger] = 1$ for $\hat{O} \in \{\hat{a}, \hat{b}\}$, and $\mathcal{N}$ includes the effects of both vacuum input noise and Brownian motion. The HLEs of motion are:
\begin{equation}
 \begin{aligned}
\dot{\hat{a}}&= -(i\Delta_{a}+\frac{\kappa}{2})\hat{a} +ig_{0}\hat{a}(\hat{b}^{\dagger}+\hat{b})+ \Omega_{c}\\ \nonumber
&+ E_{p}e^{-i\delta_{p} t} +\sqrt{2\kappa_a}\,\hat{a}_{\mathrm{in}},\\ 
\dot{\hat{b}}&= -(i\omega_{b} - \frac{\gamma}{2})\hat{b}+ig_{0}\hat{a}^{\dagger}\hat{a} +\sqrt{2\gamma}\,\hat{b}_{\mathrm{in}} .\label{SCHLEs-1}
\end{aligned} 
\end{equation}
Here, $\kappa$ and $\gamma$ denote the total decay rate and mechanical gain, respectively, while the $\hat{a}_{\mathrm{in}}$ and $\hat{b}_{\mathrm{in}}$ represent the corresponding noise operators.  Focusing on the system's mean response to the applied probe field, we neglect quantum input and thermal noise. The corresponding noise operators have zero mean and satisfy the correlation functions~\cite{Anwar-2020, Muhib-2019}:
\begin{align}
\langle \hat{a}_{\mathrm{in}}^{\dagger}(t)\hat{a}_{\mathrm{in}}(t') \rangle &= 0, 
\qquad
\langle \hat{a}_{\mathrm{in}}(t)\hat{a}_{\mathrm{in}}^{\dagger}(t') \rangle = \delta(t-t'), \nonumber\\
\langle \hat{b}_{\mathrm{in}}^{\dagger}(t)\hat{b}_{\mathrm{in}}(t') \rangle &= 0, 
\qquad
\langle \hat{b}_{\mathrm{in}}(t)\hat{b}_{\mathrm{in}}^{\dagger}(t') \rangle = \delta(t-t'). \nonumber\\
\end{align}
Replacing each operator $\hat{O}(t)$ with its expectation value $O(t) \equiv \langle \hat{O}(t) \rangle$ for $O \in \{a, b\}$~\cite{Xiong2015}, leads to:
\begin{equation}
 \begin{aligned}
\dot{a}&= -({i}\Delta_{a}+\frac{\kappa}{2})a + ig_{0}a(b^{\ast} + b)+ \Omega_{c} + E_{p}e^{-i\delta_{p} t},\\   
\dot{b}&= -(i\omega_{b} - \frac{\gamma}{2})b +ig_{0}a^{\ast}a.\label{HLEs-2}
\end{aligned} 
\end{equation}
We analyze the linear response to a weak probe field ($E_p \ll \Omega_c$), expressing each variable as $O = {O}_{s} + \delta{O} \quad (O = a, b)$, where $O_s$ is the steady state value and $\delta{O}$ the first-order fluctuation. The steady-state values are:
\begin{equation}
\begin{aligned}
a_s = \frac{\Omega_{c}}{i\Delta + \frac{\kappa}{2}}, \\
b_s = \frac{ig_0|a_{s}|^2}{i\omega_{b} - \frac{\gamma}{2}},
\end{aligned}\label{steady-states} 
\end{equation}
where $\Delta = \Delta_{a} - g_0(b^\ast_{s} + b_{s})$ is the effective detuning. The first-order fluctuations are:
\begin{equation}
\begin{aligned}
\delta\dot{a}& = -\left(i\Delta + \frac{\kappa}{2}\right)\delta{a} + iG_{ab}(\delta{b^{\ast}} + \delta{b}) + E_{p}e^{-i\delta_{p} t}\\
\delta\dot{b} &= -\left(i\omega_{b} - \frac{\gamma}{2}\right)\delta{b} + i(G_{ab}\delta{a^{\ast}}+ G_{ab}^{\ast}\delta{a}),\label{Ist-order-1}
\end{aligned}
\end{equation}
where $G_{ab} = g_{0}a_{s}$ is the effective optomechanical coupling strength. We consider the optical cavity driven in the red-sideband regime ($\Delta = \omega_b$). In the resolved-sideband limit ($\omega_b \gg \kappa, \gamma$), the rotating-wave approximation \cite{Stefan2010, Massel2012, Liu_2017} reduces Eq.~(\ref{Ist-order-1}) to
\begin{equation}
\begin{aligned}
\delta\dot{a} &= -\left(i\Delta + \frac{\kappa}{2}\right)\delta{a} + iG_{ab}\delta{b} + E_{p}e^{-i\delta_{p} t} \\
\delta\dot{b} &= -\left(i\omega_{b} -\frac{\gamma}{2}\right)\delta{b} + i G_{ab}^{\ast}\delta{a}.
\end{aligned}
\end{equation}
To simplify the dynamical analysis, we transform the system into a frame rotating at the probe frequency $\delta_{p}$, under which the first-order fluctuations evolve as $\delta{a}\rightarrow A e^{-i\delta_{p} t}$ and $\delta {b} \rightarrow B e^{-i\delta_{p} t}$,
\begin{equation}
\begin{aligned}
\dot{A} &= -\left(i\omega_1 +\frac{\kappa}{2}\right)A + iG_{ab}B + E_{p}, \\
\dot{B} &= -\left (i\omega_2 -\frac{\gamma}{2}\right)B + iG_{ab}^\ast A,
\end{aligned}\label{Ist-order-2}
\end{equation}
where $\omega_1 = \Delta - \delta_{p}$ and $\omega_2 = \omega_{b} - \delta_{p}$.
\subsection{ Effective Hamiltonian and $\mathcal{PT}$-symmetry}\label{SecIIB}
\begin{figure}
\centering
\includegraphics[width=0.9\linewidth]{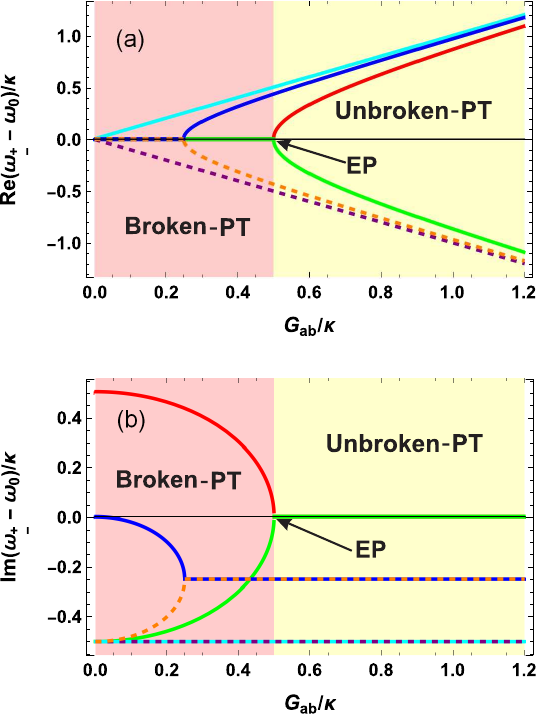}
\caption{Eigenfrequencies of $H_{\text{eff}}$ [Eq.~(\ref{H_eff})] versus normalized effective optomechanical coupling $G_{ab}/\kappa$. (a) Real part: $\mathrm{Re}(\omega_\pm-\omega_{0})/\kappa$; (b) Imaginary part: $\mathrm{Im}(\omega_\pm-\omega_{0})/\kappa$. Red and green solid curves represent the $\mathcal{PT}$-symmetric case with balanced gain and loss ($\gamma/2\pi=\kappa/2\pi=1.0~\mathrm{MHz}$). For a conventional passive–passive system, blue solid and orange dashed curves indicate unequal losses ($\gamma/2\pi=1.0~\mathrm{kHz}$, $\kappa/2\pi=1.0~\mathrm{MHz}$), while cyan solid and purple dashed curves correspond to equal losses ($\gamma/2\pi=1.0~\mathrm{MHz}$, $\kappa/2\pi=1.0~\mathrm{MHz}$).}
\label{fig2}
\end{figure}
The first order linearized HLEs [Eq.~(\ref{Ist-order-2})] can be succinctly formulated in matrix form:
\begin{equation}
\dot{\mathbf{\Psi}} = -i H_{\text{eff}}\,\mathbf{\Psi},
\end{equation}
where $\mathbf{\Psi}= (A, B)^\mathsf{T}$ is the column vector, and $H_{\text{eff}}$ represents the effective Hamiltonian of the system,
\begin{equation}
H_{\text{eff}}=
{\begin{pmatrix}
\omega_{1} -i\frac{\kappa}{2} & -G_{ab} \\
-G_{ab}^\ast & \omega_{2} +i\frac{\gamma}{2} 
\end{pmatrix}}.\label{H_eff}
\end{equation}
As evident from $H_{\text{eff}}$ [Eq.~(\ref{H_eff})], the system exhibits $\mathcal{PT}$ symmetry when the two resonators are degenerate ($\omega_1 = \omega_2 = \omega_{0}$) and the loss–gain rates are balanced ($\kappa=\gamma$). Diagonalizing the matrix in Eq.~(\ref{H_eff}), then yields the eigenfrequencies:
\begin{equation}
\omega_{\pm} = \omega_0 - \frac{i}{4}(\kappa - \gamma) \pm \sqrt{|{G_{ab}|^2 - \left(\frac{\kappa + \gamma}{4}\right)^2}}. 
\end{equation}
Figures~\ref{fig2}(a) and (b) display the eigenfrequency spectrum of the $\mathcal{PT}$-symmetric cavity optomechanical system (red and green solid curves) versus normalized effective optomechanical coupling $G_{ab}/\kappa$. Under balanced gain-loss conditions ($\gamma=\kappa$), the system exhibits a broken $\mathcal{PT}$-symmetric phase for $G_{ab} / \kappa<0.5$, characterized by a complex-conjugate eigenfrequency pair. At $G_{ab}/\kappa=0.5$, the system undergoes a $\mathcal{PT}$ phase transition, with eigenfrequencies coalescing at the $\mathrm{EP}$. For $G_{ab}/\kappa>0.5$, the system enters the unbroken $\mathcal{PT}$-symmetric phase, where the eigenfrequencies are real. 

In contrast, for a conventional passive–passive optomechanical system, two cases are considered. First, in the equal-loss case, the mechanical damping ($\gamma/2\pi=1.0~\mathrm{MHz}$) is increased to match the cavity decay rate ($\kappa/2\pi=1.0~\mathrm{MHz}$), and no phase transition occurs (cyan and purple dashed curves). Second, in the more general unequal-loss case, with mechanical damping $\gamma/2\pi=1.0~\mathrm{kHz}$ and cavity decay rate $\kappa/2\pi=1.0~\mathrm{MHz}$, the eigenfrequencies are not purely real (blue solid and orange dashed curves), as shown in Figs.~\ref{fig2}(a) and (b).

\subsection{Optical Susceptibility}
To determine the optical susceptibility of the coupled system, we solve Eq.~(\ref{Ist-order-2}) under steady-state conditions ($\dot{A}_{s}=\dot{B}_{s}=0$). This yields the steady-state intracavity field of the probe: 
\begin{equation}
A_{s} = \frac{\left(i\omega_2 - \frac{\gamma}{2}\right)E_{p}}{\left(i\omega_1 + \frac{\kappa}{2}\right)\left(i\omega_2 - \frac{\gamma}{2}\right) + |G_{ab}|^2},\label{A} 
\end{equation}
The input-output relation is expressed as $a_{\text{out}} = a_{\text{in}} - \sqrt{\kappa_{e}} \, a(t),$ where $a_{\text{in}}$ ($a_{\text{out}}$) represents the amplitude of the input (output) probe field~\cite{Ling2018, Stefan2010, Yin2025}. The optical susceptibility $\chi$ is defined via the output field $E_{\text{T}}$ of the weak probe:
\begin{equation}
\chi \equiv E_{T} = \sqrt{\kappa_e} A_{s} / E_{p},\label{optical-sus}
\end{equation}
where the $\chi = \chi_{r} + i\chi_{i}$ is the complex susceptibility, and its quadratures can be measured via homodyne detection~\cite{walls1994quantum}. The real part $\chi_{r}$ describes the probe field absorption, while the imaginary part $\chi_{i}$ characterizes its dispersion spectrum. The effective intracavity permittivity $\epsilon_{2}$ describes the cavity’s response to the probe field and is connected to the optical susceptibility through $\epsilon_{2} = 1 + \chi$.  

We consider the effective detuning $\Delta$ between the cavity and control field to the mechanical frequency $\Delta = \omega_b$. So, $\omega_1 = \omega_2 = \omega_b-\delta_{p} = -\tilde{\Delta}$, where $\tilde{\Delta}$ is the probe–cavity detuning~\cite{Liu_2017}. The Eq.~(\ref{optical-sus}) takes the form:
\begin{equation}
\chi \equiv E_{T}= \frac{\sqrt{\eta\kappa}}{\left(-i\tilde{\Delta} + \frac{\kappa}{2} \right) +\frac{|G_{ab}|^{2}}{\left(-i\tilde{\Delta} - \frac{\gamma}{2} \right)}}.\label{output-probe}
\end{equation}
\subsection{Goos–H\"{a}nchen shift}
When the transverse electric (TE) polarized probe field $E_{p}$ reflects from mirror $M_{1}$, it acquires a lateral displacement along the $z$-axis, known as the GHS $S_{r}$ (Fig.~\ref{fig1}). To determine this shift, we employ the stationary-phase method, which models a well-collimated probe beam with a narrow angular spectrum as an effective plane wave. The lateral displacement of the reflected probe field is expressed as~\cite{Artmann-1948, Li-2003PRL}
\begin{equation}
S_{r} = -\frac{\lambda}{2\pi} \frac{d\phi_{r}}{d\theta_{i}},
\end{equation}
where $\lambda$ is the probe field wavelength, $\phi_{r}$ is the phase of the TE-polarized reflection coefficient $R(k_{z}, \omega_{p})$, $k_{z} = (2\pi / \lambda) \sin{\theta_{i}}$ is the wavenumber along $z$-axis, and $\theta_{i}$ is the incidence angle. The GHS is explicitly written as~\cite{Wang2005, Wang_2008}
\begin{equation}
\begin{split}
S_{r} = -\frac{\lambda}{2\pi} \frac{1}{|R(k_{z},\omega_{p})|^2}
\left\{\operatorname{Re}[R(k_{z},\omega_{p})] \frac{d \operatorname{Im}[R(k_{z}, \omega_{p})]}{d\theta_{i}}\right. \\
\left. - \operatorname{Im}[R(k_{z}, \omega_{p})] \frac{d \operatorname{Re}[R(k_{z}, \omega_{p})]}{d\theta_{i}}
\right\}.
\label{GHS} 
\end{split}
\end{equation}
The reflection coefficient $R(k_{z}, \omega_{p})$ defined in Eq.~(\ref{GHS}) is determined via the standard transfer matrix~\cite{Wang_2008}
\begin{equation}
R(k_{z}, \omega_{p}) = \frac{q_{0}(X_{22} - X_{11}) - (q_{0}^2 X_{12} - X_{21})}{q_{0}(X_{22} + X_{11}) - (q_{0}^2 X_{12} + X_{21})}, \label{2x2}
\end{equation}
where $q_{0} = \sqrt{\epsilon_{0} - \sin^2{\theta_{i}}}$, and $X_{ij}$ (with $i,j = 1,2$) are the elements of the total transfer matrix $X(k_{z}, \omega_{p})$. The total transfer matrix of the three-layer optomechanical system is given by~\cite{Wang_2008}
\begin{widetext}
\begin{equation}   
X(k_{z}, \omega_{p}) = M_{1}(k_{z}, \omega_{p}, d_{1}) M_{2}(k_{z}, \omega_{p}, d_{2}) M_{1}(k_{z}, \omega_{p}, d_{1}) =\begin{pmatrix}
X_{11} & X_{12} \\
X_{21} & X_{22}
\end{pmatrix},
\end{equation}
\end{widetext}
where $M_{j}(k_{z}, \omega_{p},d_{j})$ depends on the parameters of the corresponding layer
\begin{equation}
M_{j}(k_{z},\omega_{p},d_{j}) =
\begin{pmatrix}
\cos[k_{x}^{j} d_j] & i \sin[k_{x}^{j} d_{j}] k/ k_{x}^{j} \\
i\sin[k_{x}^{j} d_{j}]k_{x}^{j}/k & \cos[k_{x}^{j} d_{j}]
\end{pmatrix}. \label{transfer matrix}
\end{equation}
Here $k = \omega_{p}/c$ is the vacuum wavenumber, $c$ is the speed of light and $k_{x}^{j} = k \sqrt{\epsilon_{j} - \sin^2{\theta_{i}}}$ is the $x$-component of the probe field wavenumber in the $j$-th layer. Each layer ($j\equiv1,2$) is characterized by its permittivity $\epsilon_{j}$ and thickness $d_{j}$. 
\begin{figure*}
\centering
\includegraphics[width=0.95\linewidth]{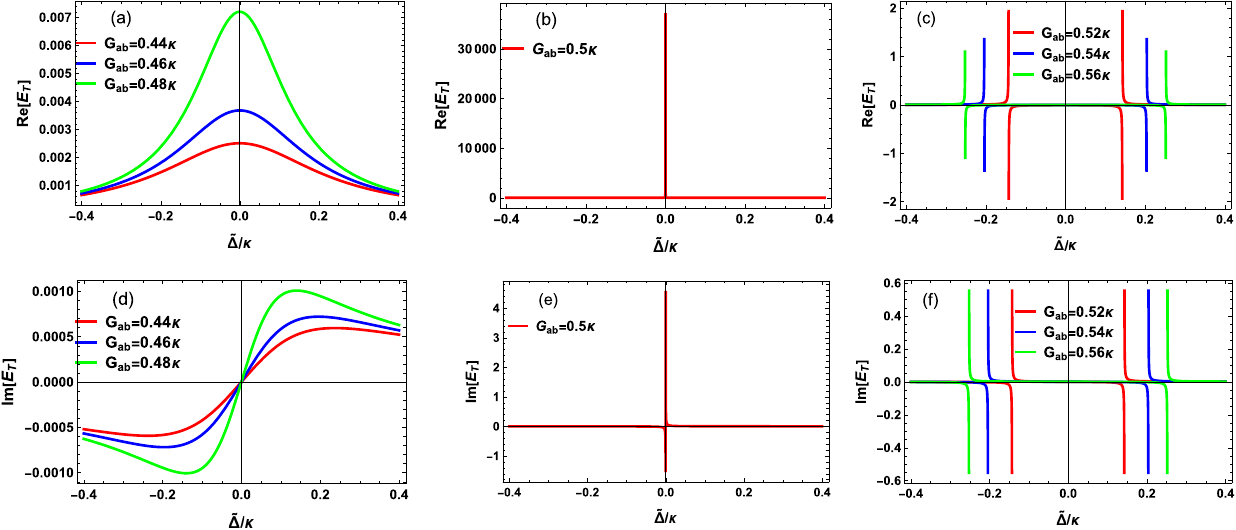}
\caption{(a–c) Absorption spectra ($\mathrm{Re}[E_T]$) and (d–f) dispersion spectra ($\mathrm{Im}[E_T]$) of the output probe field versus normalized probe–cavity detuning $\tilde{\Delta}/\kappa$.  
Columns correspond to effective optomechanical coupling strengths $G_{ab}$ for each phase: Broken $\mathcal{PT}$ phase: $G_{ab} = 0.44\,\kappa$ (red), $0.46\,\kappa$ (blue), $0.48\,\kappa$ (green); EP: $G_{ab} = 0.5\,\kappa$ (red);  
Unbroken $\mathcal{PT}$ phase: $G_{ab} = 0.52\,\kappa$ (red), $0.54\,\kappa$ (blue), $0.56\,\kappa$ (green). Fixed parameters: $\eta = 0.5$, and $\gamma/2\pi=\kappa/2\pi=1.0~\mathrm{MHz}$.}
\label{fig3}
\end{figure*}
\begin{figure}
\centering
\includegraphics[width=0.95\linewidth]{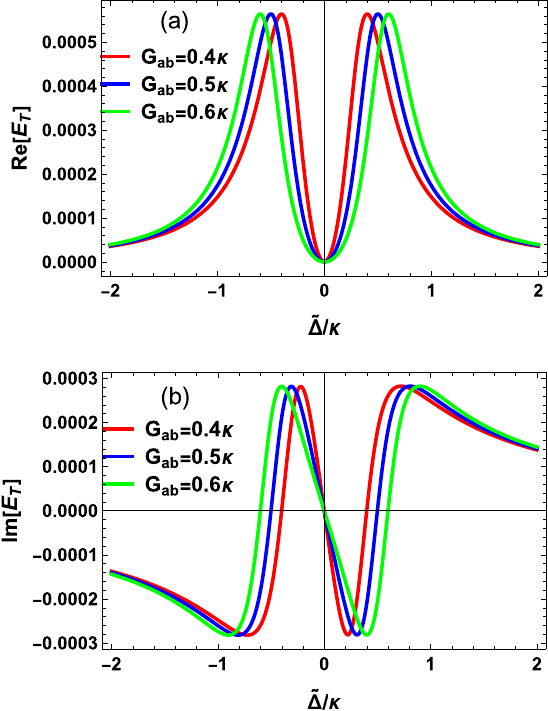}
\caption{(a) Absorption spectra ($\mathrm{Re}[E_T]$) and (b) dispersion spectra ($\mathrm{Im}[E_T]$) versus normalized probe–cavity detuning $\tilde{\Delta}/\kappa$ for various effective optomechanical coupling strengths: $G_{ab} = 0.4\,\kappa$ (red), $0.5\,\kappa$ (blue), and $0.6\,\kappa$ (green) in a conventional optomechanical system. Fixed parameters: $\eta=0.5$, $\gamma/2\pi= 1.0~\mathrm{kHz}$, and $\kappa/2\pi=1.0~\mathrm{MHz}$.}
\label{fig4}
\end{figure}
\begin{figure}[t]
\centering
\includegraphics[width=0.94\linewidth]{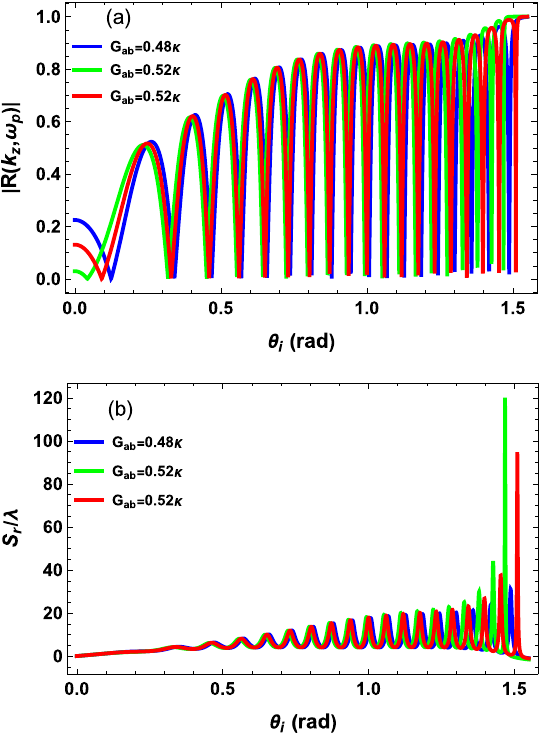}
\caption{(a) Absolute value of the reflection coefficient $|R(k_{z}, \omega_{p})|$ and (b) the normalized GHS $S_{r}/\lambda$ versus incident angle of the probe field $\theta_{i}$ for three effective optomechanical coupling strengths: (i) $G_{ab} = 0.48\,\kappa$ (blue, broken $\mathcal{PT}$ phase), (ii) $0.52\,\kappa$ (green, unbroken $\mathcal{PT}$ phase), and (iii) $0.52\,\kappa$ (red, conventional case) at resonance ($\tilde{\Delta}=0$). Fixed parameters: $\lambda=1064~\mathrm{nm}$, $\kappa/2\pi=1.0~\mathrm{MHz}$, $\epsilon_{0}=1$, $\epsilon_{1}=2.22$, $d_{1}=1~\mu\mathrm{m}$, $d_{2}=10~\mu\mathrm{m}$, $\gamma/2\pi=\kappa/2\pi=1.0~\mathrm{MHz}$ ($\mathcal{PT}$-symmetric system), $\gamma/2\pi=1.0~\mathrm{kHz}$ and $\kappa/2\pi=1.0~\mathrm{MHz}$ (conventional system), and $\eta=0.5$.}
\label{fig5}
\end{figure}
\section{Results and Discussion}
We use the following experimental accessible parameters to illustrate the manipulation of the GHS in our system~\cite{Antoni2011, Arcizet2006}: $\lambda=1064~\mathrm{nm}$, $\kappa/2\pi=1.0~\mathrm{MHz}$, $\epsilon_{0}=1$, $\epsilon_{1}=2.22$, $d_{1}=1~\mu\mathrm{m}$, $d_{2}=10~\mu\mathrm{m}$, and $\eta=0.5$. For the $\mathcal{PT}$-symmetric case, we consider $\gamma/2\pi=\kappa/2\pi=1.0~\mathrm{MHz}$, while for the conventional case, $\gamma/2\pi=1.0~\mathrm{kHz}$ and $\kappa/2\pi=1.0~\mathrm{MHz}$. Numerical results for the output probe field $E_T$ [Eq.~(\ref{output-probe})] are presented for three operating regimes: (i) broken $\mathcal{PT}$-symmetric phase, (ii) $\mathrm{EP}$, and (iii) unbroken $\mathcal{PT}$-symmetric phase, as defined in Sec.~\ref{SecIIB} [Fig.~\ref{fig2}]. Figure~\ref{fig3} depicts the probe field's absorption [$\text{Re}(E_T)$] and dispersion [$\text{Im}(E_T)$] spectra as a function of normalized probe–cavity detuning $\tilde{\Delta}/\kappa$ across these regimes. The absorption spectra are shown in panels (a)-(c). In the broken $\mathcal{PT}$-symmetric phase [panel (a)], characterized by a complex-conjugate eigenfrequency pair, the spectrum exhibits Lorentzian peaks at the resonance condition ($\tilde{\Delta}=0$). At the $\mathrm{EP}$ [panel (b)], the coalescence of eigenfrequencies results in a sharp absorption peak at $\tilde{\Delta}=0$—a direct signature of the enhanced sensitivity to perturbations characteristic of this non-Hermitian singularity. In contrast, the unbroken $\mathcal{PT}$-symmetric phase [panel (c)] reveals symmetric pairs of absorption resonances with balanced positive/negative amplitudes about $\tilde{\Delta}=0$~\cite{Ruter2010, Ziauddin2015, fahad2025PSHE}, indicative of strong effective optomechanical coupling ($G_{ab}$). The corresponding dispersion spectra are presented in panels (d)-(f). The $\mathrm{EP}$ [panel (e)] features a divergence at $\tilde{\Delta}=0$, corresponding to an abrupt $\pi$-phase shift of the probe field. In the unbroken $\mathcal{PT}$-symmetric regime [panel (f)], the dispersion spectrum displays asymmetric resonances at $\tilde{\Delta}=0$, a hallmark of $\mathcal{PT}$-symmetric systems~\cite{Ruter2010, Ziauddin2015, fahad2025PSHE}.

Figure~\ref{fig4}(a) and (b) present the output absorption spectra ($\text{Re}[E_T]$) and dispersion spectra ($\text{Im}[E_T]$) versus normalized probe–cavity detuning $\tilde{\Delta}/\kappa$ for various effective optomechanical coupling strengths ($G_{ab}$) in a conventional optomechanical system. The absorption spectra exhibit the characteristic signature of OMIT: a transparency window centered at $\tilde{\Delta}=0$ flanked by two Lorentzian-like sidebands~\cite{Stefan2010, Anwar-2020}. The width of this transparency window depends directly on $G_{ab}$, becoming progressively wider as $G_{ab}$ increases from $0.4\,\kappa$ (red) to $0.5\,\kappa$ (blue) and $0.6\,\kappa$ (green). Figure~\ref{fig4}(b) shows the corresponding dispersion spectra.

The GHS $S_{r}$ [Eq.~(\ref{GHS})] depends on the reflection coefficient $|R(k_{z}, \omega_{p})|$ [Eq.~(\ref{2x2})] of the incident TE-polarized probe field. Therefore, a detailed understanding of this reflection coefficient is essential, as it is the underlying quantity that determines the differences in the GHS across distinct $\mathcal{PT}$-symmetric phases and in comparison with conventional (non-$\mathcal{PT}$) systems. Figures~\ref{fig5}(a) and (b) show both $|R(k_{z}, \omega_{p})|$ and $S_{r}/\lambda$ as functions of the probe field incident angle $\theta_{i}$ for broken $\mathcal{PT}$ phase (blue), unbroken $\mathcal{PT}$ phase (green), and in conventional case (red) at resonance ($\tilde{\Delta}=0$). Figure~\ref{fig5}(a) reveals distinct reflection dips at specific incident angles, corresponding to resonance conditions in each $\mathcal{PT}$-symmetric phase and in the conventional system. These resonant features indicate the presence of lateral shifts, and the peaks in the GHS [Fig.~\ref{fig5}(b)] occur precisely at these reflection resonances. In the unbroken $\mathcal{PT}$ phase, the GHS is significantly enhanced, exhibiting a large positive shift. In contrast, the broken $\mathcal{PT}$ phase yields much smaller positive shifts. This difference arises from the stronger effective optomechanical coupling in the unbroken phase, which produces sharper reflection resonances and steeper phase gradients, thereby amplifying the GHS. Such behavior aligns with earlier findings in atomic-ensemble-based $\mathcal{PT}$-symmetric cavities~\cite{Ziauddin2015} and cavity magnomechanical systems~\cite{fahad2025}. At the $\mathrm{EP}$, the coalescence of eigenvalues and eigenvectors suppresses phase dispersion entirely, causing the GHS to approach zero. Therefore, we restrict our analysis to the physically meaningful broken and unbroken $\mathcal{PT}$-phases. In a conventional case, the GHS is smaller than that in the unbroken $\mathcal{PT}$-phase, demonstrating that the interplay of balanced gain–loss and strong optomechanical coupling in the unbroken $\mathcal{PT}$-phase enhances the system’s sensitivity to phase variations.

\begin{figure}
\begin{center}
\includegraphics[width=0.95\linewidth]{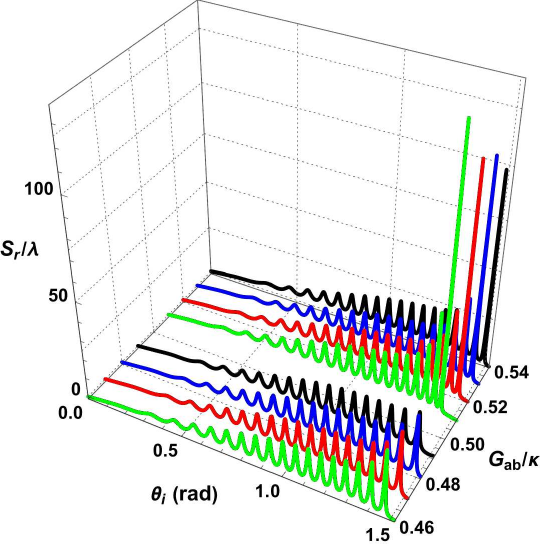}
\caption{Normalized Goos–H\"{a}nchen shift $S_r/\lambda$ as a function of the incident probe field angle $\theta_i$, parametrized by the effective optomechanical coupling strength $G_{ab}$. The curves are shown for the two dynamical regimes—broken and unbroken $\mathcal{PT}$ phases—at resonance ($\tilde{\Delta}=0$). The effective coupling strengths are: $G_{ab}/\kappa = 0.46$, $0.47$, $0.48$ $0.49$ (green, red, blue, and black curves, respectively) for the broken $\mathcal{PT}$ phase, and $0.51$, $0.52$, $0.53$, $0.54$ (green, red, blue, and black curves, respectively) for the unbroken $\mathcal{PT}$ phase. Fixed parameters: $\lambda=1064~\mathrm{nm}$, $\gamma/2\pi=\kappa/2\pi=1.0~\mathrm{MHz}$, $\epsilon_{0}=1$, $\epsilon_{1}=2.22$, $d_{1}=1~\mu\mathrm{m}$, $d_{2}=10~\mu\mathrm{m}$, and $\eta=0.5$.}
\label{fig6}
\end{center}
\end{figure}
\begin{figure}
\begin{center}
\includegraphics[width=0.95\linewidth]{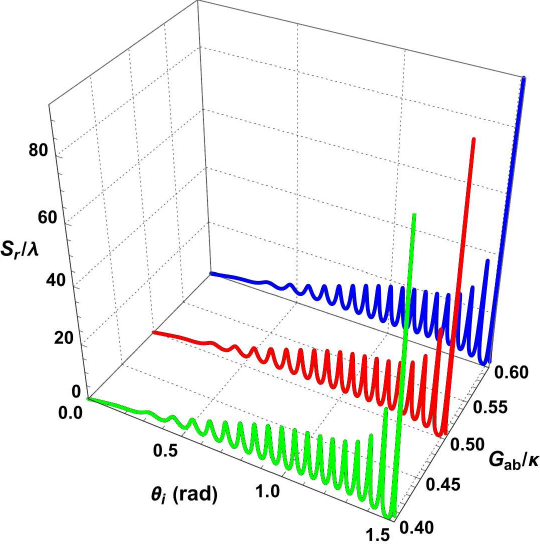}
\caption{Normalized Goos–H\"{a}nchen shift $S_r/\lambda$ as a function of the incident probe field angle $\theta_i$, parametrized by the effective optomechanical coupling strength $G_{ab}$. The curves correspond to the conventional system at resonance ($\tilde{\Delta}=0$). The effective coupling strengths are $G_{ab}/\kappa = 0.40$ (green), $0.50$ (red), and $0.60$ (blue). Fixed parameters: $\lambda=1064~\mathrm{nm}$, $\kappa/2\pi=1.0~\mathrm{MHz}$, $\epsilon_{0}=1$, $\epsilon_{1}=2.22$, $d_{1}=1~\mu\mathrm{m}$, $d_{2}=10~\mu\mathrm{m}$, $\gamma=1.0~\mathrm{kHz}$, $\kappa=1.0~\mathrm{MHz}$, and $\eta=0.5$.}
\label{fig7}
\end{center}
\end{figure}
\begin{figure*}
\begin{center}
\includegraphics[width=0.95\linewidth]{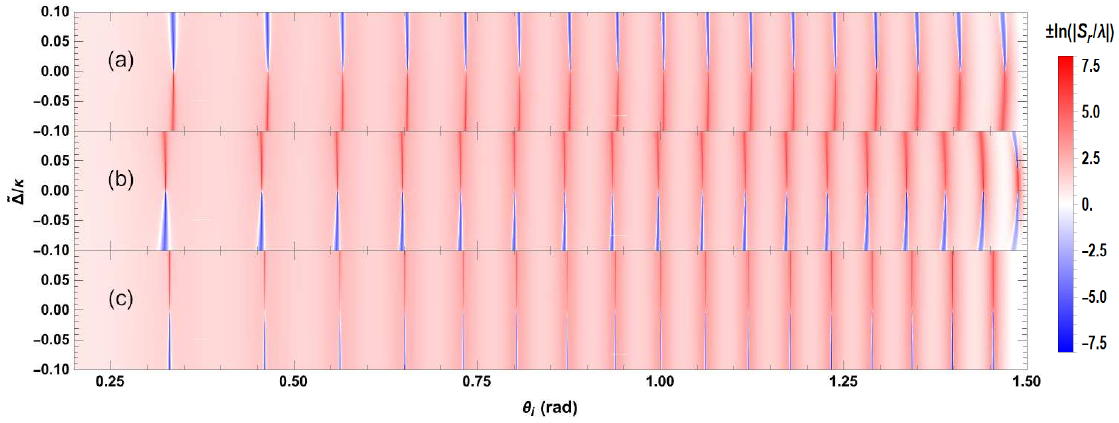}
\caption{Normalized Goos–H\"{a}nchen shift $S_r/\lambda$ versus the incident angle of the probe field $\theta_i$ and normalized probe–cavity detuning $\tilde{\Delta}/\kappa$ for (a) broken $\mathcal{PT}$ phase ($G_{ab} = 0.46\,\kappa$), (b) unbroken $\mathcal{PT}$ phase ($G_{ab} = 0.54\,\kappa$), and (c) conventional system ($G_{ab} = 0.54\,\kappa$). Fixed parameters: $\lambda = 1064~\mathrm{nm}$, $\kappa/2\pi = 1.0~\mathrm{MHz}$, $\epsilon_0 = 1$, $\epsilon_1 = 2.22$, $d_1 = 1~\mu\mathrm{m}$, $d_2 = 10~\mu\mathrm{m}$, $\gamma/2\pi=\kappa/2\pi=1.0~\mathrm{MHz}$ ($\mathcal{PT}$-symmetric system), $\gamma/2\pi = 1.0~\mathrm{kHz}$, $\kappa/2\pi=1.0~\mathrm{MHz}$ (conventional system), and $\eta = 0.5$.}
\label{fig8}
\end{center}
\end{figure*}
In Figs.~\ref{fig5}(a) and (b), we present representative values of the effective optomechanical coupling strength $G_{ab}$ for the broken and unbroken $\mathcal{PT}$-symmetric phases. To assess the generality of these observations, we extend our analysis over a wider range of $G_{ab}$ in each regime. Across this extended parameter space, the qualitative behavior remains consistent: the GHS exhibits pronounced enhancement in the unbroken $\mathcal{PT}$ phase and suppression in the broken phase [Fig.~\ref{fig6}]. Furthermore, the contrast between the broken and unbroken $\mathcal{PT}$ phases becomes more pronounced near the $\mathrm{EP}$. This consistent behavior confirms the robustness and the pivotal role of $\mathcal{PT}$ symmetry in modulating the GHS.

To further elucidate the influence of $G_{ab}$ on the GHS in the conventional configuration, we extend our analysis over a broader parameter range. As shown in Fig.~\ref{fig7}, the GHS magnitude remains essentially constant, while its angular position changes with $G_{ab}$, indicating that variations in $G_{ab}$ affect only the GHS position. By contrast, in the $\mathcal{PT}$-symmetric system under balanced gain–loss conditions, increasing $G_{ab}$ drives a phase transition from the broken to the unbroken $\mathcal{PT}$ phase, leading to pronounced changes in both the magnitude and profile of the GHS [Fig.~\ref{fig6}]. This behavior reflects the absence of non-Hermitian degeneracies in the conventional system, where variations in $G_{ab}$ shift the resonance condition without enhancing phase dispersion.
\begin{figure*}
\begin{center}
\includegraphics[width=5.6cm]{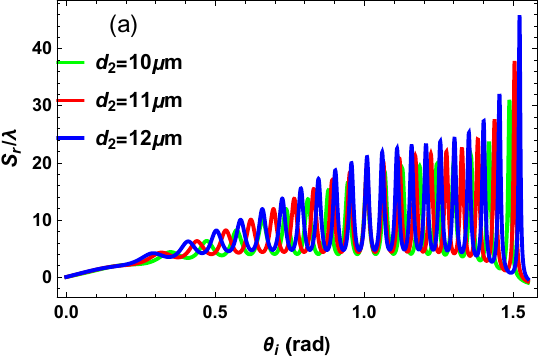}
\includegraphics[width=5.6cm]{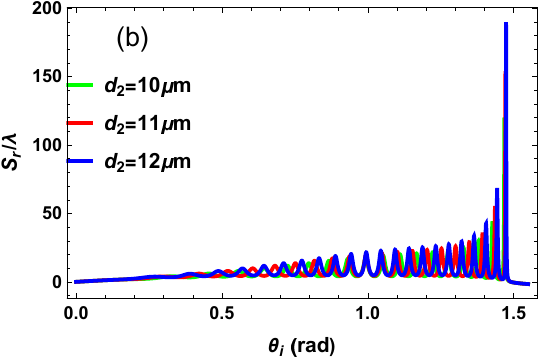}
\includegraphics[width=5.6cm]{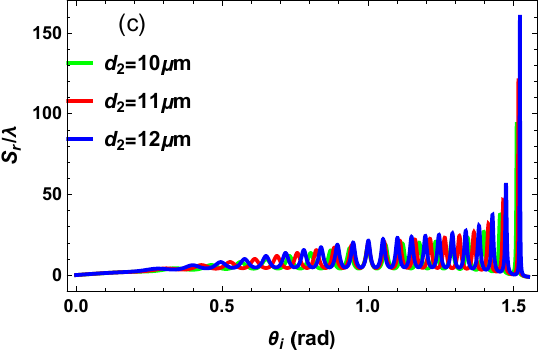}
\caption{Normalized Goos–H\"{a}nchen shift $S_{r}/\lambda$ plotted against the incident angle of the probe field $\theta_{i}$ for (a) broken $\mathcal{PT}$ phase ($G_{ab} = 0.48\,\kappa$), (b) unbroken $\mathcal{PT}$ phase ($G_{ab} = 0.52\,\kappa$), and (c) the conventional system ($G_{ab} = 0.52\,\kappa$). Each panel shows the GHS for three distinct intracavity lengths. The green, red, and blue curves correspond to $d_{2}=10$, $11$, and $12~\mu\mathrm{m}$, respectively, under resonance ($\tilde{\Delta}=0$). Fixed parameters: $\lambda=1064~\mathrm{nm}$, $\kappa/2\pi=1.0~\mathrm{MHz}$, $\epsilon_{0}=1$, $\epsilon_{1}=2.22$, $d_{1}=1~\mu\mathrm{m}$, $\gamma/2\pi=\kappa/2\pi=1.0~\mathrm{MHz}$ ($\mathcal{PT}$-symmetric system), $\gamma/2\pi=1.0~\mathrm{kHz}$, $\kappa/2\pi=1.0~\mathrm{MHz}$ (conventional system), and $\eta=0.5$.}
\label{fig9}
\end{center}
\end{figure*}

We next examine the contour plots of the GHS $S_{r}/\lambda$ against the incident angle of the probe field $\theta_{i}$ and the normalized probe–cavity detuning $\tilde{\Delta}/\kappa$ for both the $\mathcal{PT}$-symmetric system and the conventional configuration. Figures~\ref{fig8}(a-c) show the $S_{r}/\lambda$ versus $\theta_{i}$ and $\tilde{\Delta}/\kappa$ for the broken $\mathcal{PT}$ phase, unbroken $\mathcal{PT}$ phase, and conventional case, respectively. In the broken phase [Fig.~\ref{fig8}(a)], the GHS displays an asymmetric pattern with respect to the detuning, exhibiting positive lateral shifts for negative detunings and negative shifts for positive detunings. In contrast, the unbroken phase [Fig.~\ref{fig8}(b)] shows symmetric GHS patterns centered around resonance ($\tilde{\Delta}=0$)~\cite{fahad2025}. For the conventional system [Fig.~\ref{fig8}(c)], the GHS also exhibits symmetric behavior, arising from standard cavity–optomechanical interactions. Overall, the contour plots indicate that broken $\mathcal{PT}$ symmetry leads to asymmetric lateral shifts due to complex frequencies, whereas the unbroken $\mathcal{PT}$ symmetry and the conventional configuration yield symmetric shifts associated with real frequencies.

The GHS $S_{r}/ \lambda$ is highly sensitive to the intracavity length $d_2$, requiring precise dimensional control. 
Figures~\ref{fig9}(a–c) show the dependence of the $S_r/\lambda$ on the incident angle of the probe field $\theta_i$ at the resonance condition ($\tilde{\Delta}=0$) for intracavity lengths $d_2 = 10$, $11$, and $12~\mu\mathrm{m}$.
In the broken $\mathcal{PT}$ phase, both the GHS magnitude and the number of resonant peaks increase with $d_2$ [Fig.~\ref{fig9}(a)]. In the unbroken $\mathcal{PT}$ phase, the GHS exhibits larger magnitudes and a stronger dependence on $d_2$ compared to the broken phase [Fig.~\ref{fig9}(b)]. A similar trend is observed in the conventional system [Fig.~\ref{fig9}(c)]. Increasing the $d_2$ effectively lengthens the optical path within the cavity system, amplifying phase accumulation and producing a steeper phase gradient, which enhances the lateral shift. This effect is particularly pronounced in the unbroken $\mathcal{PT}$-symmetric phase due to the interplay of balanced gain–loss and stronger optomechanical coupling. The observed dependence of the GHS on intracavity length in $\mathcal{PT}$-symmetric cavity optomechanics aligns with earlier observations in cavity magnomechanics~\cite{fahad2025}, emphasizing the critical role of cavity dimensions in controlling lateral shifts.

\section{Conclusion}
In this work, we have theoretically investigated the coherent control of the GHS in both $\mathcal{PT}$-symmetric and conventional optomechanical systems. The proposed configuration, comprising a passive optical cavity coupled to an active mechanical resonator, exhibits an exceptional point (EP) at a critical effective optomechanical coupling under balanced gain-loss conditions. This EP demarcates the transition between the broken and unbroken $\mathcal{PT}$-symmetric phases.

By analyzing the absorption and dispersion spectra of the output probe field, we systematically compared the resulting GHS in the broken and unbroken $\mathcal{PT}$ phases with that in the conventional setup. Under resonant conditions ($\tilde{\Delta}=0$), the unbroken $\mathcal{PT}$-symmetric phase exhibits a significantly larger GHS compared to both the broken phase and the passive-passive configuration, emphasizing the enhancement arising from balanced gain and loss and strong effective optomechanical coupling.

We further showed that cavity detuning provides a tunable handle for controlling the lateral shift, leading to asymmetric GHS behavior in the broken phase and symmetric behavior in the unbroken and conventional cases. Variation in the length of the intracavity medium provides an additional mechanism for GHS control. Overall, our results establish a general framework for manipulating beam shifts in optomechanical systems and demonstrate that $\mathcal{PT}$-symmetric configurations enable enhanced and actively tunable photonic responses, paving the way toward reconfigurable microwave-optical devices and precision sensing applications.

\section*{Acknowledgement}
We acknowledge the financial support from the NSFC under Grant No. 12174346, and sincerely thank Muhammad Waseem for valuable discussions and insightful comments.\\
\\
\section*{references}

\bibliography{ref} 

\begin{thebibliography}{87}%
\makeatletter
\providecommand \@ifxundefined [1]{%
 \@ifx{#1\undefined}
}%
\providecommand \@ifnum [1]{%
 \ifnum #1\expandafter \@firstoftwo
 \else \expandafter \@secondoftwo
 \fi
}%
\providecommand \@ifx [1]{%
 \ifx #1\expandafter \@firstoftwo
 \else \expandafter \@secondoftwo
 \fi
}%
\providecommand \natexlab [1]{#1}%
\providecommand \enquote  [1]{``#1''}%
\providecommand \bibnamefont  [1]{#1}%
\providecommand \bibfnamefont [1]{#1}%
\providecommand \citenamefont [1]{#1}%
\providecommand \href@noop [0]{\@secondoftwo}%
\providecommand \href [0]{\begingroup \@sanitize@url \@href}%
\providecommand \@href[1]{\@@startlink{#1}\@@href}%
\providecommand \@@href[1]{\endgroup#1\@@endlink}%
\providecommand \@sanitize@url [0]{\catcode `\\12\catcode `\$12\catcode
  `\&12\catcode `\#12\catcode `\^12\catcode `\_12\catcode `\%12\relax}%
\providecommand \@@startlink[1]{}%
\providecommand \@@endlink[0]{}%
\providecommand \url  [0]{\begingroup\@sanitize@url \@url }%
\providecommand \@url [1]{\endgroup\@href {#1}{\urlprefix }}%
\providecommand \urlprefix  [0]{URL }%
\providecommand \Eprint [0]{\href }%
\providecommand \doibase [0]{http://dx.doi.org/}%
\providecommand \selectlanguage [0]{\@gobble}%
\providecommand \bibinfo  [0]{\@secondoftwo}%
\providecommand \bibfield  [0]{\@secondoftwo}%
\providecommand \translation [1]{[#1]}%
\providecommand \BibitemOpen [0]{}%
\providecommand \bibitemStop [0]{}%
\providecommand \bibitemNoStop [0]{.\EOS\space}%
\providecommand \EOS [0]{\spacefactor3000\relax}%
\providecommand \BibitemShut  [1]{\csname bibitem#1\endcsname}%
\let\auto@bib@innerbib\@empty
\bibitem [{\citenamefont {Aspelmeyer}\ \emph
  {et~al.}(2014{\natexlab{a}})\citenamefont {Aspelmeyer}, \citenamefont
  {Kippenberg},\ and\ \citenamefont {Marquardt}}]{Aspelmeyer2014}%
  \BibitemOpen
  \bibfield  {author} {\bibinfo {author} {\bibfnamefont {Markus}\ \bibnamefont
  {Aspelmeyer}}, \bibinfo {author} {\bibfnamefont {Tobias~J.}\ \bibnamefont
  {Kippenberg}}, \ and\ \bibinfo {author} {\bibfnamefont {Florian}\
  \bibnamefont {Marquardt}},\ }\bibfield  {title} {\enquote {\bibinfo {title}
  {Cavity optomechanics},}\ }\href {\doibase 10.1103/RevModPhys.86.1391}
  {\bibfield  {journal} {\bibinfo  {journal} {Rev. Mod. Phys.}\ }\textbf
  {\bibinfo {volume} {86}},\ \bibinfo {pages} {1391} (\bibinfo {year}
  {2014}{\natexlab{a}})}\BibitemShut {NoStop}%
\bibitem [{\citenamefont {Aspelmeyer}\ \emph
  {et~al.}(2014{\natexlab{b}})\citenamefont {Aspelmeyer}, \citenamefont
  {Kippenberg},\ and\ \citenamefont {Marquardt}}]{Aspelmeyer-book-2014}%
  \BibitemOpen
  \bibfield  {author} {\bibinfo {author} {\bibfnamefont {Markus}\ \bibnamefont
  {Aspelmeyer}}, \bibinfo {author} {\bibfnamefont {Tobias~J.}\ \bibnamefont
  {Kippenberg}}, \ and\ \bibinfo {author} {\bibfnamefont {Florian}\
  \bibnamefont {Marquardt}},\ }\href@noop {} {\emph {\bibinfo {title} {Cavity
  Optomechanics: Nano and Micromechanical Resonators Interacting with Light}}}\
  (\bibinfo  {publisher} {Springer},\ \bibinfo {address} {Berlin, Heidelberg},\
  \bibinfo {year} {2014})\BibitemShut {NoStop}%
\bibitem [{\citenamefont {Kippenberg}\ and\ \citenamefont
  {Vahala}(2008)}]{Kippenberg}%
  \BibitemOpen
  \bibfield  {author} {\bibinfo {author} {\bibfnamefont {T.~J.}\ \bibnamefont
  {Kippenberg}}\ and\ \bibinfo {author} {\bibfnamefont {K.~J.}\ \bibnamefont
  {Vahala}},\ }\bibfield  {title} {\enquote {\bibinfo {title} {Cavity
  optomechanics: Back-action at the mesoscale},}\ }\href {\doibase
  10.1126/science.1156032} {\bibfield  {journal} {\bibinfo  {journal}
  {Science}\ }\textbf {\bibinfo {volume} {321}},\ \bibinfo {pages} {1172}
  (\bibinfo {year} {2008})}\BibitemShut {NoStop}%
\bibitem [{\citenamefont {Purdy}\ \emph {et~al.}(2013)\citenamefont {Purdy},
  \citenamefont {Yu}, \citenamefont {Peterson}, \citenamefont {Kampel},\ and\
  \citenamefont {Regal}}]{PurdyPRX}%
  \BibitemOpen
  \bibfield  {author} {\bibinfo {author} {\bibfnamefont {T.~P.}\ \bibnamefont
  {Purdy}}, \bibinfo {author} {\bibfnamefont {P.-L.}\ \bibnamefont {Yu}},
  \bibinfo {author} {\bibfnamefont {R.~W.}\ \bibnamefont {Peterson}}, \bibinfo
  {author} {\bibfnamefont {N.~S.}\ \bibnamefont {Kampel}}, \ and\ \bibinfo
  {author} {\bibfnamefont {C.~A.}\ \bibnamefont {Regal}},\ }\bibfield  {title}
  {\enquote {\bibinfo {title} {Strong optomechanical squeezing of light},}\
  }\href {\doibase 10.1103/PhysRevX.3.031012} {\bibfield  {journal} {\bibinfo
  {journal} {Phys. Rev. X}\ }\textbf {\bibinfo {volume} {3}},\ \bibinfo {pages}
  {031012} (\bibinfo {year} {2013})}\BibitemShut {NoStop}%
\bibitem [{\citenamefont {Brooks}\ \emph {et~al.}(2012)\citenamefont {Brooks},
  \citenamefont {Botter}, \citenamefont {Schreppler}, \citenamefont {Purdy},
  \citenamefont {Brahms},\ and\ \citenamefont {Stamper-Kurn}}]{Brooks2012}%
  \BibitemOpen
  \bibfield  {author} {\bibinfo {author} {\bibfnamefont {Daniel W.~C.}\
  \bibnamefont {Brooks}}, \bibinfo {author} {\bibfnamefont {Thierry}\
  \bibnamefont {Botter}}, \bibinfo {author} {\bibfnamefont {Sydney}\
  \bibnamefont {Schreppler}}, \bibinfo {author} {\bibfnamefont {Thomas~P.}\
  \bibnamefont {Purdy}}, \bibinfo {author} {\bibfnamefont {Nathan}\
  \bibnamefont {Brahms}}, \ and\ \bibinfo {author} {\bibfnamefont {Dan~M.}\
  \bibnamefont {Stamper-Kurn}},\ }\bibfield  {title} {\enquote {\bibinfo
  {title} {Non-classical light generated by quantum-noise-driven cavity
  optomechanics},}\ }\href {\doibase 10.1038/nature11325} {\bibfield  {journal}
  {\bibinfo  {journal} {Nature}\ }\textbf {\bibinfo {volume} {488}},\ \bibinfo
  {pages} {476} (\bibinfo {year} {2012})}\BibitemShut {NoStop}%
\bibitem [{\citenamefont {Chan}\ \emph {et~al.}(2011)\citenamefont {Chan},
  \citenamefont {Alegre}, \citenamefont {Safavi-Naeini}, \citenamefont {Hill},
  \citenamefont {Krause}, \citenamefont {Gr{\"o}blacher}, \citenamefont
  {Aspelmeyer},\ and\ \citenamefont {Painter}}]{Chan2011}%
  \BibitemOpen
  \bibfield  {author} {\bibinfo {author} {\bibfnamefont {Jasper}\ \bibnamefont
  {Chan}}, \bibinfo {author} {\bibfnamefont {T.~P.~Mayer}\ \bibnamefont
  {Alegre}}, \bibinfo {author} {\bibfnamefont {Amir~H.}\ \bibnamefont
  {Safavi-Naeini}}, \bibinfo {author} {\bibfnamefont {Jeff~T.}\ \bibnamefont
  {Hill}}, \bibinfo {author} {\bibfnamefont {Alex}\ \bibnamefont {Krause}},
  \bibinfo {author} {\bibfnamefont {Simon}\ \bibnamefont {Gr{\"o}blacher}},
  \bibinfo {author} {\bibfnamefont {Markus}\ \bibnamefont {Aspelmeyer}}, \ and\
  \bibinfo {author} {\bibfnamefont {Oskar}\ \bibnamefont {Painter}},\
  }\bibfield  {title} {\enquote {\bibinfo {title} {Laser cooling of a
  nanomechanical oscillator into its quantum ground state},}\ }\href {\doibase
  10.1038/nature10461} {\bibfield  {journal} {\bibinfo  {journal} {Nature}\
  }\textbf {\bibinfo {volume} {478}},\ \bibinfo {pages} {89} (\bibinfo {year}
  {2011})}\BibitemShut {NoStop}%
\bibitem [{\citenamefont {Pontin}\ \emph {et~al.}(2018)\citenamefont {Pontin},
  \citenamefont {Bonaldi}, \citenamefont {Borrielli}, \citenamefont {Marconi},
  \citenamefont {Marino}, \citenamefont {Pandraud}, \citenamefont {Prodi},
  \citenamefont {Sarro}, \citenamefont {Serra},\ and\ \citenamefont
  {Marin}}]{PontinPRA}%
  \BibitemOpen
  \bibfield  {author} {\bibinfo {author} {\bibfnamefont {A.}~\bibnamefont
  {Pontin}}, \bibinfo {author} {\bibfnamefont {M.}~\bibnamefont {Bonaldi}},
  \bibinfo {author} {\bibfnamefont {A.}~\bibnamefont {Borrielli}}, \bibinfo
  {author} {\bibfnamefont {L.}~\bibnamefont {Marconi}}, \bibinfo {author}
  {\bibfnamefont {F.}~\bibnamefont {Marino}}, \bibinfo {author} {\bibfnamefont
  {G.}~\bibnamefont {Pandraud}}, \bibinfo {author} {\bibfnamefont {G.~A.}\
  \bibnamefont {Prodi}}, \bibinfo {author} {\bibfnamefont {P.~M.}\ \bibnamefont
  {Sarro}}, \bibinfo {author} {\bibfnamefont {E.}~\bibnamefont {Serra}}, \ and\
  \bibinfo {author} {\bibfnamefont {F.}~\bibnamefont {Marin}},\ }\bibfield
  {title} {\enquote {\bibinfo {title} {Quantum nondemolition measurement of
  optical field fluctuations by optomechanical interaction},}\ }\href {\doibase
  10.1103/PhysRevA.97.033833} {\bibfield  {journal} {\bibinfo  {journal} {Phys.
  Rev. A}\ }\textbf {\bibinfo {volume} {97}},\ \bibinfo {pages} {033833}
  (\bibinfo {year} {2018})}\BibitemShut {NoStop}%
\bibitem [{\citenamefont {Massel}\ \emph {et~al.}(2011)\citenamefont {Massel},
  \citenamefont {Heikkilä}, \citenamefont {Pirkkalainen}, \citenamefont {Cho},
  \citenamefont {Saloniemi}, \citenamefont {Hakonen},\ and\ \citenamefont
  {Sillanpää}}]{Massel2011}%
  \BibitemOpen
  \bibfield  {author} {\bibinfo {author} {\bibfnamefont {Francesco}\
  \bibnamefont {Massel}}, \bibinfo {author} {\bibfnamefont {T.~T.}\
  \bibnamefont {Heikkilä}}, \bibinfo {author} {\bibfnamefont {J.-M.}\
  \bibnamefont {Pirkkalainen}}, \bibinfo {author} {\bibfnamefont {S.~U.}\
  \bibnamefont {Cho}}, \bibinfo {author} {\bibfnamefont {H.}~\bibnamefont
  {Saloniemi}}, \bibinfo {author} {\bibfnamefont {P.~J.}\ \bibnamefont
  {Hakonen}}, \ and\ \bibinfo {author} {\bibfnamefont {M.~A.}\ \bibnamefont
  {Sillanpää}},\ }\bibfield  {title} {\enquote {\bibinfo {title} {Microwave
  amplification with nanomechanical resonators},}\ }\href {\doibase
  10.1038/nature10628} {\bibfield  {journal} {\bibinfo  {journal} {Nature}\
  }\textbf {\bibinfo {volume} {480}},\ \bibinfo {pages} {351} (\bibinfo {year}
  {2011})}\BibitemShut {NoStop}%
\bibitem [{\citenamefont {{Zhou}}\ \emph {et~al.}(2013)\citenamefont {{Zhou}},
  \citenamefont {{Hocke}}, \citenamefont {{Schliesser}}, \citenamefont
  {{Marx}}, \citenamefont {{Huebl}}, \citenamefont {{Gross}},\ and\
  \citenamefont {{Kippenberg}}}]{Zhou2013}%
  \BibitemOpen
  \bibfield  {author} {\bibinfo {author} {\bibfnamefont {X.}~\bibnamefont
  {{Zhou}}}, \bibinfo {author} {\bibfnamefont {F.}~\bibnamefont {{Hocke}}},
  \bibinfo {author} {\bibfnamefont {A.}~\bibnamefont {{Schliesser}}}, \bibinfo
  {author} {\bibfnamefont {A.}~\bibnamefont {{Marx}}}, \bibinfo {author}
  {\bibfnamefont {H.}~\bibnamefont {{Huebl}}}, \bibinfo {author} {\bibfnamefont
  {R.}~\bibnamefont {{Gross}}}, \ and\ \bibinfo {author} {\bibfnamefont
  {T.~J.}\ \bibnamefont {{Kippenberg}}},\ }\bibfield  {title} {\enquote
  {\bibinfo {title} {{Slowing, advancing and switching of microwave signals
  using circuit nanoelectromechanics}},}\ }\href {\doibase 10.1038/nphys2527}
  {\bibfield  {journal} {\bibinfo  {journal} {Nat. Phys.}\ }\textbf {\bibinfo
  {volume} {9}},\ \bibinfo {pages} {179} (\bibinfo {year} {2013})}\BibitemShut
  {NoStop}%
\bibitem [{\citenamefont {Monifi}\ \emph {et~al.}(2016)\citenamefont {Monifi},
  \citenamefont {Zhang}, \citenamefont {{\"{O}}zdemir}, \citenamefont {Peng},
  \citenamefont {Liu}, \citenamefont {Bo}, \citenamefont {Nori},\ and\
  \citenamefont {Yang}}]{Monifi2016}%
  \BibitemOpen
  \bibfield  {author} {\bibinfo {author} {\bibfnamefont {Faraz}\ \bibnamefont
  {Monifi}}, \bibinfo {author} {\bibfnamefont {Jing}\ \bibnamefont {Zhang}},
  \bibinfo {author} {\bibfnamefont {{\c{S}}ahin~Kaya}\ \bibnamefont
  {{\"{O}}zdemir}}, \bibinfo {author} {\bibfnamefont {Bo}~\bibnamefont {Peng}},
  \bibinfo {author} {\bibfnamefont {Yu-Xi}\ \bibnamefont {Liu}}, \bibinfo
  {author} {\bibfnamefont {Fang}\ \bibnamefont {Bo}}, \bibinfo {author}
  {\bibfnamefont {Franco}\ \bibnamefont {Nori}}, \ and\ \bibinfo {author}
  {\bibfnamefont {Lan}\ \bibnamefont {Yang}},\ }\bibfield  {title} {\enquote
  {\bibinfo {title} {Optomechanically induced stochastic resonance and chaos
  transfer between optical fields},}\ }\href {\doibase 10.1038/nphoton.2016.73}
  {\bibfield  {journal} {\bibinfo  {journal} {Nat. Photonics}\ }\textbf
  {\bibinfo {volume} {10}},\ \bibinfo {pages} {399} (\bibinfo {year}
  {2016})}\BibitemShut {NoStop}%
\bibitem [{\citenamefont {Massel}\ \emph {et~al.}(2012)\citenamefont {Massel},
  \citenamefont {Cho}, \citenamefont {Pirkkalainen}, \citenamefont {Hakonen},
  \citenamefont {Heikkil{\"a}},\ and\ \citenamefont
  {Sillanp{\"a}{\"a}}}]{Massel2012}%
  \BibitemOpen
  \bibfield  {author} {\bibinfo {author} {\bibfnamefont {Francesco}\
  \bibnamefont {Massel}}, \bibinfo {author} {\bibfnamefont {Sung~Un}\
  \bibnamefont {Cho}}, \bibinfo {author} {\bibfnamefont {Juha-Matti}\
  \bibnamefont {Pirkkalainen}}, \bibinfo {author} {\bibfnamefont
  {Pertti~Juhani}\ \bibnamefont {Hakonen}}, \bibinfo {author} {\bibfnamefont
  {Tero~T.}\ \bibnamefont {Heikkil{\"a}}}, \ and\ \bibinfo {author}
  {\bibfnamefont {Mika~A.}\ \bibnamefont {Sillanp{\"a}{\"a}}},\ }\bibfield
  {title} {\enquote {\bibinfo {title} {Multimode circuit optomechanics near the
  quantum limit},}\ }\href {https://api.semanticscholar.org/CorpusID:5460885}
  {\bibfield  {journal} {\bibinfo  {journal} {Nat. Commun.}\ }\textbf {\bibinfo
  {volume} {3}},\ \bibinfo {pages} {987} (\bibinfo {year} {2012})}\BibitemShut
  {NoStop}%
\bibitem [{\citenamefont {Weis}\ \emph {et~al.}(2010)\citenamefont {Weis},
  \citenamefont {Rivi\`{e}re}, \citenamefont {Del\'{e}glise}, \citenamefont
  {Gavartin}, \citenamefont {Arcizet}, \citenamefont {Schliesser},\ and\
  \citenamefont {Kippenberg}}]{Stefan2010}%
  \BibitemOpen
  \bibfield  {author} {\bibinfo {author} {\bibfnamefont {Stefan}\ \bibnamefont
  {Weis}}, \bibinfo {author} {\bibfnamefont {R\'{e}mi}\ \bibnamefont
  {Rivi\`{e}re}}, \bibinfo {author} {\bibfnamefont {Samuel}\ \bibnamefont
  {Del\'{e}glise}}, \bibinfo {author} {\bibfnamefont {Emanuel}\ \bibnamefont
  {Gavartin}}, \bibinfo {author} {\bibfnamefont {Olivier}\ \bibnamefont
  {Arcizet}}, \bibinfo {author} {\bibfnamefont {Albert}\ \bibnamefont
  {Schliesser}}, \ and\ \bibinfo {author} {\bibfnamefont {Tobias~J.}\
  \bibnamefont {Kippenberg}},\ }\bibfield  {title} {\enquote {\bibinfo {title}
  {Optomechanically induced transparency},}\ }\href {\doibase
  10.1126/science.1195596} {\bibfield  {journal} {\bibinfo  {journal}
  {Science}\ }\textbf {\bibinfo {volume} {330}},\ \bibinfo {pages} {1520}
  (\bibinfo {year} {2010})}\BibitemShut {NoStop}%
\bibitem [{\citenamefont {Karuza}\ \emph {et~al.}(2013)\citenamefont {Karuza},
  \citenamefont {Biancofiore}, \citenamefont {Bawaj}, \citenamefont
  {Molinelli}, \citenamefont {Galassi}, \citenamefont {Natali}, \citenamefont
  {Tombesi}, \citenamefont {Di~Giuseppe},\ and\ \citenamefont
  {Vitali}}]{Karuza2013}%
  \BibitemOpen
  \bibfield  {author} {\bibinfo {author} {\bibfnamefont {M.}~\bibnamefont
  {Karuza}}, \bibinfo {author} {\bibfnamefont {C.}~\bibnamefont {Biancofiore}},
  \bibinfo {author} {\bibfnamefont {M.}~\bibnamefont {Bawaj}}, \bibinfo
  {author} {\bibfnamefont {C.}~\bibnamefont {Molinelli}}, \bibinfo {author}
  {\bibfnamefont {M.}~\bibnamefont {Galassi}}, \bibinfo {author} {\bibfnamefont
  {R.}~\bibnamefont {Natali}}, \bibinfo {author} {\bibfnamefont
  {P.}~\bibnamefont {Tombesi}}, \bibinfo {author} {\bibfnamefont
  {G.}~\bibnamefont {Di~Giuseppe}}, \ and\ \bibinfo {author} {\bibfnamefont
  {D.}~\bibnamefont {Vitali}},\ }\bibfield  {title} {\enquote {\bibinfo {title}
  {Optomechanically induced transparency in a membrane-in-the-middle setup at
  room temperature},}\ }\href {\doibase 10.1103/PhysRevA.88.013804} {\bibfield
  {journal} {\bibinfo  {journal} {Phys. Rev. A}\ }\textbf {\bibinfo {volume}
  {88}},\ \bibinfo {pages} {013804} (\bibinfo {year} {2013})}\BibitemShut
  {NoStop}%
\bibitem [{\citenamefont {Bender}\ and\ \citenamefont
  {Boettcher}(1998)}]{BenderPRL}%
  \BibitemOpen
  \bibfield  {author} {\bibinfo {author} {\bibfnamefont {Carl~M.}\ \bibnamefont
  {Bender}}\ and\ \bibinfo {author} {\bibfnamefont {Stefan}\ \bibnamefont
  {Boettcher}},\ }\bibfield  {title} {\enquote {\bibinfo {title} {Real spectra
  in non-hermitian hamiltonians having $\mathcal{PT}$ symmetry},}\ }\href
  {\doibase 10.1103/PhysRevLett.80.5243} {\bibfield  {journal} {\bibinfo
  {journal} {Phys. Rev. Lett.}\ }\textbf {\bibinfo {volume} {80}},\ \bibinfo
  {pages} {5243} (\bibinfo {year} {1998})}\BibitemShut {NoStop}%
\bibitem [{\citenamefont {Chtchelkatchev}\ \emph {et~al.}(2012)\citenamefont
  {Chtchelkatchev}, \citenamefont {Golubov}, \citenamefont {Baturina},\ and\
  \citenamefont {Vinokur}}]{Chtchelkatchev-PRL}%
  \BibitemOpen
  \bibfield  {author} {\bibinfo {author} {\bibfnamefont {N.~M.}\ \bibnamefont
  {Chtchelkatchev}}, \bibinfo {author} {\bibfnamefont {A.~A.}\ \bibnamefont
  {Golubov}}, \bibinfo {author} {\bibfnamefont {T.~I.}\ \bibnamefont
  {Baturina}}, \ and\ \bibinfo {author} {\bibfnamefont {V.~M.}\ \bibnamefont
  {Vinokur}},\ }\bibfield  {title} {\enquote {\bibinfo {title} {Stimulation of
  the fluctuation superconductivity by $\mathcal{PT}$ symmetry},}\ }\href
  {\doibase 10.1103/PhysRevLett.109.150405} {\bibfield  {journal} {\bibinfo
  {journal} {Phys. Rev. Lett.}\ }\textbf {\bibinfo {volume} {109}},\ \bibinfo
  {pages} {150405} (\bibinfo {year} {2012})}\BibitemShut {NoStop}%
\bibitem [{\citenamefont {Bittner}\ \emph {et~al.}(2012)\citenamefont
  {Bittner}, \citenamefont {Dietz}, \citenamefont {G{\"{u}}nther},
  \citenamefont {Harney}, \citenamefont {Miski-Oglu}, \citenamefont {Richter},\
  and\ \citenamefont {Sch{\"{a}}fer}}]{Bittner2012}%
  \BibitemOpen
  \bibfield  {author} {\bibinfo {author} {\bibfnamefont {S.}~\bibnamefont
  {Bittner}}, \bibinfo {author} {\bibfnamefont {B.}~\bibnamefont {Dietz}},
  \bibinfo {author} {\bibfnamefont {U.}~\bibnamefont {G{\"{u}}nther}}, \bibinfo
  {author} {\bibfnamefont {H.~L.}\ \bibnamefont {Harney}}, \bibinfo {author}
  {\bibfnamefont {M.}~\bibnamefont {Miski-Oglu}}, \bibinfo {author}
  {\bibfnamefont {A.}~\bibnamefont {Richter}}, \ and\ \bibinfo {author}
  {\bibfnamefont {F.}~\bibnamefont {Sch{\"{a}}fer}},\ }\bibfield  {title}
  {\enquote {\bibinfo {title} {$\mathcal{PT}$ symmetry and spontaneous symmetry
  breaking in a microwave billiard},}\ }\href {\doibase
  10.1103/PhysRevLett.108.024101} {\bibfield  {journal} {\bibinfo  {journal}
  {Phys. Rev. Lett.}\ }\textbf {\bibinfo {volume} {108}},\ \bibinfo {pages}
  {024101} (\bibinfo {year} {2012})}\BibitemShut {NoStop}%
\bibitem [{\citenamefont {Bender}\ \emph
  {et~al.}(2013{\natexlab{a}})\citenamefont {Bender}, \citenamefont {Berntson},
  \citenamefont {Parker},\ and\ \citenamefont {Samuel}}]{Bender2013}%
  \BibitemOpen
  \bibfield  {author} {\bibinfo {author} {\bibfnamefont {Carl~M.}\ \bibnamefont
  {Bender}}, \bibinfo {author} {\bibfnamefont {Bjorn~K.}\ \bibnamefont
  {Berntson}}, \bibinfo {author} {\bibfnamefont {David}\ \bibnamefont
  {Parker}}, \ and\ \bibinfo {author} {\bibfnamefont {E.}~\bibnamefont
  {Samuel}},\ }\bibfield  {title} {\enquote {\bibinfo {title} {Observation of
  $\mathcal{PT}$ phase transition in a simple mechanical system},}\ }\href
  {\doibase 10.1119/1.4789549} {\bibfield  {journal} {\bibinfo  {journal} {Am.
  J. Phys.}\ }\textbf {\bibinfo {volume} {81}},\ \bibinfo {pages} {173}
  (\bibinfo {year} {2013}{\natexlab{a}})}\BibitemShut {NoStop}%
\bibitem [{\citenamefont {Shi}\ \emph {et~al.}(2016)\citenamefont {Shi},
  \citenamefont {Dubois}, \citenamefont {Chen}, \citenamefont {Cheng},
  \citenamefont {Ramezani}, \citenamefont {Wang},\ and\ \citenamefont
  {Zhang}}]{Shi2016}%
  \BibitemOpen
  \bibfield  {author} {\bibinfo {author} {\bibfnamefont {Chengzhi}\
  \bibnamefont {Shi}}, \bibinfo {author} {\bibfnamefont {Marc}\ \bibnamefont
  {Dubois}}, \bibinfo {author} {\bibfnamefont {Yun}\ \bibnamefont {Chen}},
  \bibinfo {author} {\bibfnamefont {Lei}\ \bibnamefont {Cheng}}, \bibinfo
  {author} {\bibfnamefont {Hamidreza}\ \bibnamefont {Ramezani}}, \bibinfo
  {author} {\bibfnamefont {Yuan}\ \bibnamefont {Wang}}, \ and\ \bibinfo
  {author} {\bibfnamefont {Xiang}\ \bibnamefont {Zhang}},\ }\bibfield  {title}
  {\enquote {\bibinfo {title} {Accessing the exceptional points of parity-time
  symmetric acoustics},}\ }\href {\doibase 10.1038/ncomms11110} {\bibfield
  {journal} {\bibinfo  {journal} {Nat. Commun}\ }\textbf {\bibinfo {volume}
  {7}},\ \bibinfo {pages} {11110} (\bibinfo {year} {2016})}\BibitemShut
  {NoStop}%
\bibitem [{\citenamefont {Ruschhaupt}\ \emph {et~al.}(2005)\citenamefont
  {Ruschhaupt}, \citenamefont {Delgado},\ and\ \citenamefont
  {Muga}}]{Ruschhaupt2005}%
  \BibitemOpen
  \bibfield  {author} {\bibinfo {author} {\bibfnamefont {A}~\bibnamefont
  {Ruschhaupt}}, \bibinfo {author} {\bibfnamefont {F}~\bibnamefont {Delgado}},
  \ and\ \bibinfo {author} {\bibfnamefont {J~G}\ \bibnamefont {Muga}},\
  }\bibfield  {title} {\enquote {\bibinfo {title} {Physical realization of
  $\mathcal{PT}$-symmetric potential scattering in a planar slab waveguide},}\
  }\href {\doibase 10.1088/0305-4470/38/9/L03} {\bibfield  {journal} {\bibinfo
  {journal} {J. Phys. A Math. Gen.}\ }\textbf {\bibinfo {volume} {38}},\
  \bibinfo {pages} {L171} (\bibinfo {year} {2005})}\BibitemShut {NoStop}%
\bibitem [{\citenamefont {Bender}\ \emph {et~al.}(2004)\citenamefont {Bender},
  \citenamefont {Brody},\ and\ \citenamefont {Jones}}]{Bender2004}%
  \BibitemOpen
  \bibfield  {author} {\bibinfo {author} {\bibfnamefont {Carl~M.}\ \bibnamefont
  {Bender}}, \bibinfo {author} {\bibfnamefont {Dorje~C.}\ \bibnamefont
  {Brody}}, \ and\ \bibinfo {author} {\bibfnamefont {Hugh~F.}\ \bibnamefont
  {Jones}},\ }\bibfield  {title} {\enquote {\bibinfo {title} {Extension of
  $\mathcal{PT}$-symmetric quantum mechanics to quantum field theory with cubic
  interaction},}\ }\href {\doibase 10.1103/PhysRevD.70.025001} {\bibfield
  {journal} {\bibinfo  {journal} {Phys. Rev. D}\ }\textbf {\bibinfo {volume}
  {70}},\ \bibinfo {pages} {025001} (\bibinfo {year} {2004})}\BibitemShut
  {NoStop}%
\bibitem [{\citenamefont {Bender}\ \emph
  {et~al.}(2013{\natexlab{b}})\citenamefont {Bender}, \citenamefont
  {Gianfreda}, \citenamefont {{\"{O}}zdemir}, \citenamefont {Peng},\ and\
  \citenamefont {Yang}}]{Bender2013TWO}%
  \BibitemOpen
  \bibfield  {author} {\bibinfo {author} {\bibfnamefont {Carl~M.}\ \bibnamefont
  {Bender}}, \bibinfo {author} {\bibfnamefont {Maria}\ \bibnamefont
  {Gianfreda}}, \bibinfo {author} {\bibfnamefont {{\c{S}}ahin~Kaya}\
  \bibnamefont {{\"{O}}zdemir}}, \bibinfo {author} {\bibfnamefont
  {Bo}~\bibnamefont {Peng}}, \ and\ \bibinfo {author} {\bibfnamefont {Lan}\
  \bibnamefont {Yang}},\ }\bibfield  {title} {\enquote {\bibinfo {title}
  {Twofold transition in $\mathcal{PT}$-symmetric coupled oscillators},}\
  }\href {\doibase 10.1103/PhysRevA.88.062111} {\bibfield  {journal} {\bibinfo
  {journal} {Phys. Rev. A}\ }\textbf {\bibinfo {volume} {88}},\ \bibinfo
  {pages} {062111} (\bibinfo {year} {2013}{\natexlab{b}})}\BibitemShut
  {NoStop}%
\bibitem [{\citenamefont {El-Ganainy}\ \emph {et~al.}(2007)\citenamefont
  {El-Ganainy}, \citenamefont {Makris}, \citenamefont {Christodoulides},\ and\
  \citenamefont {Musslimani}}]{El2007theory}%
  \BibitemOpen
  \bibfield  {author} {\bibinfo {author} {\bibfnamefont {Ramy}\ \bibnamefont
  {El-Ganainy}}, \bibinfo {author} {\bibfnamefont {KG}~\bibnamefont {Makris}},
  \bibinfo {author} {\bibfnamefont {DN}~\bibnamefont {Christodoulides}}, \ and\
  \bibinfo {author} {\bibfnamefont {Ziad~H}\ \bibnamefont {Musslimani}},\
  }\bibfield  {title} {\enquote {\bibinfo {title} {Theory of coupled optical
  $\mathcal{PT}$-symmetric structures},}\ }\href {\doibase
  10.1364/OL.32.002632} {\bibfield  {journal} {\bibinfo  {journal} {Opt.
  Lett.}\ }\textbf {\bibinfo {volume} {32}},\ \bibinfo {pages} {2632} (\bibinfo
  {year} {2007})}\BibitemShut {NoStop}%
\bibitem [{\citenamefont {Iorsh}\ \emph {et~al.}(2020)\citenamefont {Iorsh},
  \citenamefont {Poshakinskiy},\ and\ \citenamefont {Poddubny}}]{Iorsh-PRL}%
  \BibitemOpen
  \bibfield  {author} {\bibinfo {author} {\bibfnamefont {Ivan}\ \bibnamefont
  {Iorsh}}, \bibinfo {author} {\bibfnamefont {Alexander}\ \bibnamefont
  {Poshakinskiy}}, \ and\ \bibinfo {author} {\bibfnamefont {Alexander}\
  \bibnamefont {Poddubny}},\ }\bibfield  {title} {\enquote {\bibinfo {title}
  {Waveguide quantum optomechanics: Parity-time phase transitions in
  ultrastrong coupling regime},}\ }\href {\doibase
  10.1103/PhysRevLett.125.183601} {\bibfield  {journal} {\bibinfo  {journal}
  {Phys. Rev. Lett.}\ }\textbf {\bibinfo {volume} {125}},\ \bibinfo {pages}
  {183601} (\bibinfo {year} {2020})}\BibitemShut {NoStop}%
\bibitem [{\citenamefont {Liu}\ \emph {et~al.}(2017)\citenamefont {Liu},
  \citenamefont {Wu}, \citenamefont {Zhang}, \citenamefont {{\"{O}}zdemir},
  \citenamefont {Yang}, \citenamefont {Nori},\ and\ \citenamefont
  {Liu}}]{Liu_2017}%
  \BibitemOpen
  \bibfield  {author} {\bibinfo {author} {\bibfnamefont {Yu-Long}\ \bibnamefont
  {Liu}}, \bibinfo {author} {\bibfnamefont {Rebing}\ \bibnamefont {Wu}},
  \bibinfo {author} {\bibfnamefont {Jing}\ \bibnamefont {Zhang}}, \bibinfo
  {author} {\bibfnamefont {{\c{S}}ahin~Kaya}\ \bibnamefont {{\"{O}}zdemir}},
  \bibinfo {author} {\bibfnamefont {Lan}\ \bibnamefont {Yang}}, \bibinfo
  {author} {\bibfnamefont {Franco}\ \bibnamefont {Nori}}, \ and\ \bibinfo
  {author} {\bibfnamefont {Yu-Xi}\ \bibnamefont {Liu}},\ }\bibfield  {title}
  {\enquote {\bibinfo {title} {Controllable optical response by modifying the
  gain and loss of a mechanical resonator and cavity mode in an optomechanical
  system},}\ }\href {\doibase 10.1103/PhysRevA.95.013843} {\bibfield  {journal}
  {\bibinfo  {journal} {Phys. Rev. A}\ }\textbf {\bibinfo {volume} {95}},\
  \bibinfo {pages} {013843} (\bibinfo {year} {2017})}\BibitemShut {NoStop}%
\bibitem [{\citenamefont {Peng}\ \emph
  {et~al.}(2014{\natexlab{a}})\citenamefont {Peng}, \citenamefont
  {{\"O}zdemir}, \citenamefont {Lei}, \citenamefont {Monifi}, \citenamefont
  {Gianfreda}, \citenamefont {Long}, \citenamefont {Fan}, \citenamefont {Nori},
  \citenamefont {Bender},\ and\ \citenamefont {Yang}}]{Peng2014}%
  \BibitemOpen
  \bibfield  {author} {\bibinfo {author} {\bibfnamefont {Bo}~\bibnamefont
  {Peng}}, \bibinfo {author} {\bibfnamefont {{\c{S}}ahin~Kaya}\ \bibnamefont
  {{\"O}zdemir}}, \bibinfo {author} {\bibfnamefont {Fuchuan}\ \bibnamefont
  {Lei}}, \bibinfo {author} {\bibfnamefont {Faraz}\ \bibnamefont {Monifi}},
  \bibinfo {author} {\bibfnamefont {Mariagiovanna}\ \bibnamefont {Gianfreda}},
  \bibinfo {author} {\bibfnamefont {Gui~Lu}\ \bibnamefont {Long}}, \bibinfo
  {author} {\bibfnamefont {Shanhui}\ \bibnamefont {Fan}}, \bibinfo {author}
  {\bibfnamefont {Franco}\ \bibnamefont {Nori}}, \bibinfo {author}
  {\bibfnamefont {Carl~M.}\ \bibnamefont {Bender}}, \ and\ \bibinfo {author}
  {\bibfnamefont {Lan}\ \bibnamefont {Yang}},\ }\bibfield  {title} {\enquote
  {\bibinfo {title} {Parity-time-symmetric whispering-gallery microcavities},}\
  }\href {\doibase 10.1038/nphys2927} {\bibfield  {journal} {\bibinfo
  {journal} {Nat. Phys.}\ }\textbf {\bibinfo {volume} {10}},\ \bibinfo {pages}
  {394} (\bibinfo {year} {2014}{\natexlab{a}})}\BibitemShut {NoStop}%
\bibitem [{\citenamefont {Peng}\ \emph
  {et~al.}(2014{\natexlab{b}})\citenamefont {Peng}, \citenamefont
  {{\"O}zdemir}, \citenamefont {Rotter}, \citenamefont {Yilmaz}, \citenamefont
  {Liertzer}, \citenamefont {Monifi}, \citenamefont {Bender}, \citenamefont
  {Nori},\ and\ \citenamefont {Yang}}]{Peng2014Science}%
  \BibitemOpen
  \bibfield  {author} {\bibinfo {author} {\bibfnamefont {B.}~\bibnamefont
  {Peng}}, \bibinfo {author} {\bibfnamefont {{\c{S}}.~K.}\ \bibnamefont
  {{\"O}zdemir}}, \bibinfo {author} {\bibfnamefont {S.}~\bibnamefont {Rotter}},
  \bibinfo {author} {\bibfnamefont {H.}~\bibnamefont {Yilmaz}}, \bibinfo
  {author} {\bibfnamefont {M.}~\bibnamefont {Liertzer}}, \bibinfo {author}
  {\bibfnamefont {F.}~\bibnamefont {Monifi}}, \bibinfo {author} {\bibfnamefont
  {C.~M.}\ \bibnamefont {Bender}}, \bibinfo {author} {\bibfnamefont
  {F.}~\bibnamefont {Nori}}, \ and\ \bibinfo {author} {\bibfnamefont
  {L.}~\bibnamefont {Yang}},\ }\bibfield  {title} {\enquote {\bibinfo {title}
  {Loss-induced suppression and revival of lasing},}\ }\href {\doibase
  10.1126/science.1258004} {\bibfield  {journal} {\bibinfo  {journal}
  {Science}\ }\textbf {\bibinfo {volume} {346}},\ \bibinfo {pages} {328}
  (\bibinfo {year} {2014}{\natexlab{b}})}\BibitemShut {NoStop}%
\bibitem [{\citenamefont {{R\"{u}}ter}\ \emph {et~al.}(2010)\citenamefont
  {{R\"{u}}ter}, \citenamefont {Makris}, \citenamefont {El-Ganainy},
  \citenamefont {Christodoulides}, \citenamefont {Segev},\ and\ \citenamefont
  {Kip}}]{Ruter2010}%
  \BibitemOpen
  \bibfield  {author} {\bibinfo {author} {\bibfnamefont {Christian~E.}\
  \bibnamefont {{R\"{u}}ter}}, \bibinfo {author} {\bibfnamefont
  {Konstantinos~G.}\ \bibnamefont {Makris}}, \bibinfo {author} {\bibfnamefont
  {Ramy}\ \bibnamefont {El-Ganainy}}, \bibinfo {author} {\bibfnamefont
  {Demetrios~N.}\ \bibnamefont {Christodoulides}}, \bibinfo {author}
  {\bibfnamefont {Mordechai}\ \bibnamefont {Segev}}, \ and\ \bibinfo {author}
  {\bibfnamefont {Detlef}\ \bibnamefont {Kip}},\ }\bibfield  {title} {\enquote
  {\bibinfo {title} {Observation of parity–time symmetry in optics},}\ }\href
  {\doibase 10.1038/nphys1515} {\bibfield  {journal} {\bibinfo  {journal} {Nat.
  Phys.}\ }\textbf {\bibinfo {volume} {6}},\ \bibinfo {pages} {192} (\bibinfo
  {year} {2010})}\BibitemShut {NoStop}%
\bibitem [{\citenamefont {Konotop}\ \emph {et~al.}(2012)\citenamefont
  {Konotop}, \citenamefont {Shchesnovich},\ and\ \citenamefont
  {Zezyulin}}]{KONOTOP2012}%
  \BibitemOpen
  \bibfield  {author} {\bibinfo {author} {\bibfnamefont {Vladimir~V.}\
  \bibnamefont {Konotop}}, \bibinfo {author} {\bibfnamefont {Valery~S.}\
  \bibnamefont {Shchesnovich}}, \ and\ \bibinfo {author} {\bibfnamefont
  {Dmitry~A.}\ \bibnamefont {Zezyulin}},\ }\bibfield  {title} {\enquote
  {\bibinfo {title} {Giant amplification of modes in parity-time symmetric
  waveguides},}\ }\href {\doibase
  https://doi.org/10.1016/j.physleta.2012.07.027} {\bibfield  {journal}
  {\bibinfo  {journal} {Phys. Lett. A}\ }\textbf {\bibinfo {volume} {376}},\
  \bibinfo {pages} {2750} (\bibinfo {year} {2012})}\BibitemShut {NoStop}%
\bibitem [{\citenamefont {Feng}\ \emph {et~al.}(2013)\citenamefont {Feng},
  \citenamefont {Xu}, \citenamefont {Fegadolli}, \citenamefont {Lu},
  \citenamefont {Oliveira}, \citenamefont {Almeida}, \citenamefont {Chen},\
  and\ \citenamefont {Scherer}}]{Feng2013}%
  \BibitemOpen
  \bibfield  {author} {\bibinfo {author} {\bibfnamefont {Liang}\ \bibnamefont
  {Feng}}, \bibinfo {author} {\bibfnamefont {Ye-Long}\ \bibnamefont {Xu}},
  \bibinfo {author} {\bibfnamefont {William~S.}\ \bibnamefont {Fegadolli}},
  \bibinfo {author} {\bibfnamefont {Ming-Hui}\ \bibnamefont {Lu}}, \bibinfo
  {author} {\bibfnamefont {José E.~B.}\ \bibnamefont {Oliveira}}, \bibinfo
  {author} {\bibfnamefont {Vilson~R.}\ \bibnamefont {Almeida}}, \bibinfo
  {author} {\bibfnamefont {Yan-Feng}\ \bibnamefont {Chen}}, \ and\ \bibinfo
  {author} {\bibfnamefont {Axel}\ \bibnamefont {Scherer}},\ }\bibfield  {title}
  {\enquote {\bibinfo {title} {Experimental demonstration of a unidirectional
  reflectionless parity-time metamaterial at optical frequencies},}\ }\href
  {\doibase 10.1038/nmat3495} {\bibfield  {journal} {\bibinfo  {journal} {Nat.
  Mat.}\ }\textbf {\bibinfo {volume} {12}},\ \bibinfo {pages} {108} (\bibinfo
  {year} {2013})}\BibitemShut {NoStop}%
\bibitem [{\citenamefont {Lin}\ \emph {et~al.}(2011)\citenamefont {Lin},
  \citenamefont {Ramezani}, \citenamefont {Eichelkraut}, \citenamefont
  {Kottos}, \citenamefont {Cao},\ and\ \citenamefont
  {Christodoulides}}]{LinPRL}%
  \BibitemOpen
  \bibfield  {author} {\bibinfo {author} {\bibfnamefont {Zin}\ \bibnamefont
  {Lin}}, \bibinfo {author} {\bibfnamefont {Hamidreza}\ \bibnamefont
  {Ramezani}}, \bibinfo {author} {\bibfnamefont {Toni}\ \bibnamefont
  {Eichelkraut}}, \bibinfo {author} {\bibfnamefont {Tsampikos}\ \bibnamefont
  {Kottos}}, \bibinfo {author} {\bibfnamefont {Hui}\ \bibnamefont {Cao}}, \
  and\ \bibinfo {author} {\bibfnamefont {Demetrios~N.}\ \bibnamefont
  {Christodoulides}},\ }\bibfield  {title} {\enquote {\bibinfo {title}
  {Unidirectional invisibility induced by $\mathcal{P}\mathcal{T}$-symmetric
  periodic structures},}\ }\href {\doibase 10.1103/PhysRevLett.106.213901}
  {\bibfield  {journal} {\bibinfo  {journal} {Phys. Rev. Lett.}\ }\textbf
  {\bibinfo {volume} {106}},\ \bibinfo {pages} {213901} (\bibinfo {year}
  {2011})}\BibitemShut {NoStop}%
\bibitem [{\citenamefont {Schindler}\ \emph {et~al.}(2011)\citenamefont
  {Schindler}, \citenamefont {Li}, \citenamefont {Zheng}, \citenamefont
  {Ellis},\ and\ \citenamefont {Kottos}}]{Schindle2011}%
  \BibitemOpen
  \bibfield  {author} {\bibinfo {author} {\bibfnamefont {Joseph}\ \bibnamefont
  {Schindler}}, \bibinfo {author} {\bibfnamefont {Ang}\ \bibnamefont {Li}},
  \bibinfo {author} {\bibfnamefont {Mei~C.}\ \bibnamefont {Zheng}}, \bibinfo
  {author} {\bibfnamefont {F.~M.}\ \bibnamefont {Ellis}}, \ and\ \bibinfo
  {author} {\bibfnamefont {Tsampikos}\ \bibnamefont {Kottos}},\ }\bibfield
  {title} {\enquote {\bibinfo {title} {Experimental study of active lrc
  circuits with $\mathcal{PT}$ symmetries},}\ }\href {\doibase
  10.1103/PhysRevA.84.040101} {\bibfield  {journal} {\bibinfo  {journal} {Phys.
  Rev. A}\ }\textbf {\bibinfo {volume} {84}},\ \bibinfo {pages} {040101}
  (\bibinfo {year} {2011})}\BibitemShut {NoStop}%
\bibitem [{\citenamefont {Jiao}\ \emph {et~al.}(2016)\citenamefont {Jiao},
  \citenamefont {Lü}, \citenamefont {Qian}, \citenamefont {Li},\ and\
  \citenamefont {Jing}}]{Jiao-2016}%
  \BibitemOpen
  \bibfield  {author} {\bibinfo {author} {\bibfnamefont {Y}~\bibnamefont
  {Jiao}}, \bibinfo {author} {\bibfnamefont {H}~\bibnamefont {Lü}}, \bibinfo
  {author} {\bibfnamefont {J}~\bibnamefont {Qian}}, \bibinfo {author}
  {\bibfnamefont {Y}~\bibnamefont {Li}}, \ and\ \bibinfo {author}
  {\bibfnamefont {H}~\bibnamefont {Jing}},\ }\bibfield  {title} {\enquote
  {\bibinfo {title} {Nonlinear optomechanics with gain and loss: amplifying
  higher-order sideband and group delay},}\ }\href {\doibase
  10.1088/1367-2630/18/8/083034} {\bibfield  {journal} {\bibinfo  {journal}
  {New J. Phys.}\ }\textbf {\bibinfo {volume} {18}},\ \bibinfo {pages} {083034}
  (\bibinfo {year} {2016})}\BibitemShut {NoStop}%
\bibitem [{\citenamefont {Li}\ \emph {et~al.}(2016)\citenamefont {Li},
  \citenamefont {Jiang}, \citenamefont {Li},\ and\ \citenamefont
  {Song}}]{Li2016}%
  \BibitemOpen
  \bibfield  {author} {\bibinfo {author} {\bibfnamefont {Wenlin}\ \bibnamefont
  {Li}}, \bibinfo {author} {\bibfnamefont {Yunfeng}\ \bibnamefont {Jiang}},
  \bibinfo {author} {\bibfnamefont {Chong}\ \bibnamefont {Li}}, \ and\ \bibinfo
  {author} {\bibfnamefont {Heshan}\ \bibnamefont {Song}},\ }\bibfield  {title}
  {\enquote {\bibinfo {title} {Parity-time-symmetry enhanced
  optomechanically-induced-transparency},}\ }\href {\doibase 10.1038/srep31095}
  {\bibfield  {journal} {\bibinfo  {journal} {Sci. Rep.}\ }\textbf {\bibinfo
  {volume} {6}},\ \bibinfo {pages} {31095} (\bibinfo {year}
  {2016})}\BibitemShut {NoStop}%
\bibitem [{\citenamefont {Liu}\ \emph {et~al.}(2016)\citenamefont {Liu},
  \citenamefont {Zhang}, \citenamefont {{\"{O}}zdemir}, \citenamefont {Peng},
  \citenamefont {Jing}, \citenamefont {{L\"{u}}}, \citenamefont {Li},
  \citenamefont {Yang}, \citenamefont {Nori},\ and\ \citenamefont
  {Liu}}]{Liu2016PRL}%
  \BibitemOpen
  \bibfield  {author} {\bibinfo {author} {\bibfnamefont {Zhong-Peng}\
  \bibnamefont {Liu}}, \bibinfo {author} {\bibfnamefont {Jing}\ \bibnamefont
  {Zhang}}, \bibinfo {author} {\bibfnamefont {{\c{S}}ahin~Kaya}\ \bibnamefont
  {{\"{O}}zdemir}}, \bibinfo {author} {\bibfnamefont {Bo}~\bibnamefont {Peng}},
  \bibinfo {author} {\bibfnamefont {Hui}\ \bibnamefont {Jing}}, \bibinfo
  {author} {\bibfnamefont {Xin-You}\ \bibnamefont {{L\"{u}}}}, \bibinfo
  {author} {\bibfnamefont {Chun-Wen}\ \bibnamefont {Li}}, \bibinfo {author}
  {\bibfnamefont {Lan}\ \bibnamefont {Yang}}, \bibinfo {author} {\bibfnamefont
  {Franco}\ \bibnamefont {Nori}}, \ and\ \bibinfo {author} {\bibfnamefont
  {Yu-Xi}\ \bibnamefont {Liu}},\ }\bibfield  {title} {\enquote {\bibinfo
  {title} {Metrology with $\mathcal{PT}$-symmetric cavities: Enhanced
  sensitivity near the $\mathcal{PT}$-phase transition},}\ }\href {\doibase
  10.1103/PhysRevLett.117.110802} {\bibfield  {journal} {\bibinfo  {journal}
  {Phys. Rev. Lett.}\ }\textbf {\bibinfo {volume} {117}},\ \bibinfo {pages}
  {110802} (\bibinfo {year} {2016})}\BibitemShut {NoStop}%
\bibitem [{\citenamefont {He}\ \emph {et~al.}(2016)\citenamefont {He},
  \citenamefont {Yang},\ and\ \citenamefont {Xiao}}]{He-2016}%
  \BibitemOpen
  \bibfield  {author} {\bibinfo {author} {\bibfnamefont {Bing}\ \bibnamefont
  {He}}, \bibinfo {author} {\bibfnamefont {Liu}\ \bibnamefont {Yang}}, \ and\
  \bibinfo {author} {\bibfnamefont {Min}\ \bibnamefont {Xiao}},\ }\bibfield
  {title} {\enquote {\bibinfo {title} {Dynamical phonon laser in coupled
  active-passive microresonators},}\ }\href {\doibase
  10.1103/PhysRevA.94.031802} {\bibfield  {journal} {\bibinfo  {journal} {Phys.
  Rev. A}\ }\textbf {\bibinfo {volume} {94}},\ \bibinfo {pages} {031802}
  (\bibinfo {year} {2016})}\BibitemShut {NoStop}%
\bibitem [{\citenamefont {{L\"{u}}}\ \emph {et~al.}(2015)\citenamefont
  {{L\"{u}}}, \citenamefont {Jing}, \citenamefont {Ma},\ and\ \citenamefont
  {Wu}}]{Lu-2015}%
  \BibitemOpen
  \bibfield  {author} {\bibinfo {author} {\bibfnamefont {Xin-You}\ \bibnamefont
  {{L\"{u}}}}, \bibinfo {author} {\bibfnamefont {Hui}\ \bibnamefont {Jing}},
  \bibinfo {author} {\bibfnamefont {Jin-Yong}\ \bibnamefont {Ma}}, \ and\
  \bibinfo {author} {\bibfnamefont {Ying}\ \bibnamefont {Wu}},\ }\bibfield
  {title} {\enquote {\bibinfo {title} {$\mathcal{PT}$-symmetry-breaking chaos
  in optomechanics},}\ }\href {\doibase 10.1103/PhysRevLett.114.253601}
  {\bibfield  {journal} {\bibinfo  {journal} {Phys. Rev. Lett.}\ }\textbf
  {\bibinfo {volume} {114}},\ \bibinfo {pages} {253601} (\bibinfo {year}
  {2015})}\BibitemShut {NoStop}%
\bibitem [{\citenamefont {He}(2019)}]{He-2019}%
  \BibitemOpen
  \bibfield  {author} {\bibinfo {author} {\bibfnamefont {Ling-Yan}\
  \bibnamefont {He}},\ }\bibfield  {title} {\enquote {\bibinfo {title}
  {Parity-time-symmetry-enhanced sideband generation in an optomechanical
  system},}\ }\href {\doibase 10.1103/PhysRevA.99.033843} {\bibfield  {journal}
  {\bibinfo  {journal} {Phys. Rev. A}\ }\textbf {\bibinfo {volume} {99}},\
  \bibinfo {pages} {033843} (\bibinfo {year} {2019})}\BibitemShut {NoStop}%
\bibitem [{\citenamefont {Jing}\ \emph {et~al.}(2015)\citenamefont {Jing},
  \citenamefont {{\"O}zdemir}, \citenamefont {Geng}, \citenamefont {Zhang},
  \citenamefont {L{\"u}}, \citenamefont {Peng}, \citenamefont {Yang},\ and\
  \citenamefont {Nori}}]{Jing2015}%
  \BibitemOpen
  \bibfield  {author} {\bibinfo {author} {\bibfnamefont {H.}~\bibnamefont
  {Jing}}, \bibinfo {author} {\bibfnamefont {{\c{S}}ahin~K.}\ \bibnamefont
  {{\"O}zdemir}}, \bibinfo {author} {\bibfnamefont {Z.}~\bibnamefont {Geng}},
  \bibinfo {author} {\bibfnamefont {Jing}\ \bibnamefont {Zhang}}, \bibinfo
  {author} {\bibfnamefont {Xin-You}\ \bibnamefont {L{\"u}}}, \bibinfo {author}
  {\bibfnamefont {Bo}~\bibnamefont {Peng}}, \bibinfo {author} {\bibfnamefont
  {Lan}\ \bibnamefont {Yang}}, \ and\ \bibinfo {author} {\bibfnamefont
  {Franco}\ \bibnamefont {Nori}},\ }\bibfield  {title} {\enquote {\bibinfo
  {title} {Optomechanically-induced transparency in parity-time-symmetric
  microresonators},}\ }\href {\doibase 10.1038/srep09663} {\bibfield  {journal}
  {\bibinfo  {journal} {Sci. Rep.}\ }\textbf {\bibinfo {volume} {5}},\ \bibinfo
  {pages} {9663} (\bibinfo {year} {2015})}\BibitemShut {NoStop}%
\bibitem [{\citenamefont {Zhang}\ \emph {et~al.}(2017)\citenamefont {Zhang},
  \citenamefont {Guo}, \citenamefont {Pei},\ and\ \citenamefont
  {Yi}}]{Zhang-2017}%
  \BibitemOpen
  \bibfield  {author} {\bibinfo {author} {\bibfnamefont {X.~Y.}\ \bibnamefont
  {Zhang}}, \bibinfo {author} {\bibfnamefont {Y.~Q.}\ \bibnamefont {Guo}},
  \bibinfo {author} {\bibfnamefont {P.}~\bibnamefont {Pei}}, \ and\ \bibinfo
  {author} {\bibfnamefont {X.~X.}\ \bibnamefont {Yi}},\ }\bibfield  {title}
  {\enquote {\bibinfo {title} {Optomechanically induced absorption in
  parity-time-symmetric optomechanical systems},}\ }\href {\doibase
  10.1103/PhysRevA.95.063825} {\bibfield  {journal} {\bibinfo  {journal} {Phys.
  Rev. A}\ }\textbf {\bibinfo {volume} {95}},\ \bibinfo {pages} {063825}
  (\bibinfo {year} {2017})}\BibitemShut {NoStop}%
\bibitem [{\citenamefont {Jing}\ \emph {et~al.}(2014)\citenamefont {Jing},
  \citenamefont {{\"{O}}zdemir}, \citenamefont {{L\"{u}}}, \citenamefont
  {Zhang}, \citenamefont {Yang},\ and\ \citenamefont {Nori}}]{Jing-PRL}%
  \BibitemOpen
  \bibfield  {author} {\bibinfo {author} {\bibfnamefont {Hui}\ \bibnamefont
  {Jing}}, \bibinfo {author} {\bibfnamefont {S.~K.}\ \bibnamefont
  {{\"{O}}zdemir}}, \bibinfo {author} {\bibfnamefont {Xin-You}\ \bibnamefont
  {{L\"{u}}}}, \bibinfo {author} {\bibfnamefont {Jing}\ \bibnamefont {Zhang}},
  \bibinfo {author} {\bibfnamefont {Lan}\ \bibnamefont {Yang}}, \ and\ \bibinfo
  {author} {\bibfnamefont {Franco}\ \bibnamefont {Nori}},\ }\bibfield  {title}
  {\enquote {\bibinfo {title} {$\mathcal{PT}$-symmetric phonon laser},}\ }\href
  {\doibase 10.1103/PhysRevLett.113.053604} {\bibfield  {journal} {\bibinfo
  {journal} {Phys. Rev. Lett.}\ }\textbf {\bibinfo {volume} {113}},\ \bibinfo
  {pages} {053604} (\bibinfo {year} {2014})}\BibitemShut {NoStop}%
\bibitem [{\citenamefont {Zhang}\ \emph {et~al.}(2018)\citenamefont {Zhang},
  \citenamefont {Peng}, \citenamefont {{\"O}zdemir}, \citenamefont {Pichler},
  \citenamefont {Krimer}, \citenamefont {Zhao}, \citenamefont {Nori},
  \citenamefont {xi~Liu}, \citenamefont {Rotter},\ and\ \citenamefont
  {Yang}}]{Zhang2018}%
  \BibitemOpen
  \bibfield  {author} {\bibinfo {author} {\bibfnamefont {Jing}\ \bibnamefont
  {Zhang}}, \bibinfo {author} {\bibfnamefont {Bo}~\bibnamefont {Peng}},
  \bibinfo {author} {\bibfnamefont {{\c{S}}ahin~Kaya}\ \bibnamefont
  {{\"O}zdemir}}, \bibinfo {author} {\bibfnamefont {Kevin}\ \bibnamefont
  {Pichler}}, \bibinfo {author} {\bibfnamefont {Dmitry~O.}\ \bibnamefont
  {Krimer}}, \bibinfo {author} {\bibfnamefont {Guangming}\ \bibnamefont
  {Zhao}}, \bibinfo {author} {\bibfnamefont {Franco}\ \bibnamefont {Nori}},
  \bibinfo {author} {\bibfnamefont {Yu}~\bibnamefont {xi~Liu}}, \bibinfo
  {author} {\bibfnamefont {Stefan}\ \bibnamefont {Rotter}}, \ and\ \bibinfo
  {author} {\bibfnamefont {Lan}\ \bibnamefont {Yang}},\ }\bibfield  {title}
  {\enquote {\bibinfo {title} {A phonon laser operating at an exceptional
  point},}\ }\href {\doibase 10.1038/s41566-018-0213-5} {\bibfield  {journal}
  {\bibinfo  {journal} {Nat. Photonics}\ }\textbf {\bibinfo {volume} {12}},\
  \bibinfo {pages} {479--484} (\bibinfo {year} {2018})}\BibitemShut {NoStop}%
\bibitem [{\citenamefont {Zhong}\ \emph {et~al.}(2019)\citenamefont {Zhong},
  \citenamefont {Nelson}, \citenamefont {\c{S}. K.~\"{O}zdemir},\ and\
  \citenamefont {El-Ganainy}}]{Zhong2019}%
  \BibitemOpen
  \bibfield  {author} {\bibinfo {author} {\bibfnamefont {Q.}~\bibnamefont
  {Zhong}}, \bibinfo {author} {\bibfnamefont {S.}~\bibnamefont {Nelson}},
  \bibinfo {author} {\bibnamefont {\c{S}. K.~\"{O}zdemir}}, \ and\ \bibinfo
  {author} {\bibfnamefont {R.}~\bibnamefont {El-Ganainy}},\ }\bibfield  {title}
  {\enquote {\bibinfo {title} {Controlling directional absorption with chiral
  exceptional surfaces},}\ }\href {\doibase 10.1364/OL.44.005242} {\bibfield
  {journal} {\bibinfo  {journal} {Opt. Lett.}\ }\textbf {\bibinfo {volume}
  {44}},\ \bibinfo {pages} {5242} (\bibinfo {year} {2019})}\BibitemShut
  {NoStop}%
\bibitem [{\citenamefont {O'Connell}\ \emph {et~al.}(2010)\citenamefont
  {O'Connell}, \citenamefont {Hofheinz}, \citenamefont {Ansmann}, \citenamefont
  {Bialczak}, \citenamefont {Lenander}, \citenamefont {Lucero}, \citenamefont
  {Neeley}, \citenamefont {Sank}, \citenamefont {Wang}, \citenamefont {Weides},
  \citenamefont {Wenner}, \citenamefont {Martinis},\ and\ \citenamefont
  {Cleland}}]{OConnell2010}%
  \BibitemOpen
  \bibfield  {author} {\bibinfo {author} {\bibfnamefont {Aaron~D.}\
  \bibnamefont {O'Connell}}, \bibinfo {author} {\bibfnamefont {Max}\
  \bibnamefont {Hofheinz}}, \bibinfo {author} {\bibfnamefont {Markus}\
  \bibnamefont {Ansmann}}, \bibinfo {author} {\bibfnamefont {Radoslaw~C.}\
  \bibnamefont {Bialczak}}, \bibinfo {author} {\bibfnamefont {Mike}\
  \bibnamefont {Lenander}}, \bibinfo {author} {\bibfnamefont {Erik}\
  \bibnamefont {Lucero}}, \bibinfo {author} {\bibfnamefont {Matthew}\
  \bibnamefont {Neeley}}, \bibinfo {author} {\bibfnamefont {Daniel~Thomas}\
  \bibnamefont {Sank}}, \bibinfo {author} {\bibfnamefont {Haohua}\ \bibnamefont
  {Wang}}, \bibinfo {author} {\bibfnamefont {Martin~P.}\ \bibnamefont
  {Weides}}, \bibinfo {author} {\bibfnamefont {James}\ \bibnamefont {Wenner}},
  \bibinfo {author} {\bibfnamefont {John~M.}\ \bibnamefont {Martinis}}, \ and\
  \bibinfo {author} {\bibfnamefont {Andrew~N.}\ \bibnamefont {Cleland}},\
  }\bibfield  {title} {\enquote {\bibinfo {title} {Quantum ground state and
  single-phonon control of a mechanical resonator},}\ }\href
  {https://api.semanticscholar.org/CorpusID:4412475} {\bibfield  {journal}
  {\bibinfo  {journal} {Nature}\ }\textbf {\bibinfo {volume} {464}},\ \bibinfo
  {pages} {697} (\bibinfo {year} {2010})}\BibitemShut {NoStop}%
\bibitem [{\citenamefont {Fan}\ \emph {et~al.}(2015)\citenamefont {Fan},
  \citenamefont {Fong}, \citenamefont {Poot},\ and\ \citenamefont
  {Tang}}]{Fan2015}%
  \BibitemOpen
  \bibfield  {author} {\bibinfo {author} {\bibfnamefont {Linran}\ \bibnamefont
  {Fan}}, \bibinfo {author} {\bibfnamefont {King~Y.}\ \bibnamefont {Fong}},
  \bibinfo {author} {\bibfnamefont {Menno}\ \bibnamefont {Poot}}, \ and\
  \bibinfo {author} {\bibfnamefont {Hong~X.}\ \bibnamefont {Tang}},\ }\bibfield
   {title} {\enquote {\bibinfo {title} {Cascaded optical transparency in
  multimode-cavity optomechanical systems},}\ }\href {\doibase
  10.1038/ncomms6850} {\bibfield  {journal} {\bibinfo  {journal} {Nat.
  Commun.}\ }\textbf {\bibinfo {volume} {6}},\ \bibinfo {pages} {5850}
  (\bibinfo {year} {2015})}\BibitemShut {NoStop}%
\bibitem [{\citenamefont {Okamoto}\ \emph {et~al.}(2013)\citenamefont
  {Okamoto}, \citenamefont {Gourgout}, \citenamefont {Chang}, \citenamefont
  {Onomitsu}, \citenamefont {Mahboob}, \citenamefont {Chang},\ and\
  \citenamefont {Yamaguchi}}]{Okamoto2013}%
  \BibitemOpen
  \bibfield  {author} {\bibinfo {author} {\bibfnamefont {Hajime}\ \bibnamefont
  {Okamoto}}, \bibinfo {author} {\bibfnamefont {Adrien}\ \bibnamefont
  {Gourgout}}, \bibinfo {author} {\bibfnamefont {Chia-Yuan}\ \bibnamefont
  {Chang}}, \bibinfo {author} {\bibfnamefont {Koji}\ \bibnamefont {Onomitsu}},
  \bibinfo {author} {\bibfnamefont {Imran}\ \bibnamefont {Mahboob}}, \bibinfo
  {author} {\bibfnamefont {Edward~Yi}\ \bibnamefont {Chang}}, \ and\ \bibinfo
  {author} {\bibfnamefont {Hiroshi}\ \bibnamefont {Yamaguchi}},\ }\bibfield
  {title} {\enquote {\bibinfo {title} {Coherent phonon manipulation in coupled
  mechanical resonators},}\ }\href {\doibase 10.1038/nphys2665} {\bibfield
  {journal} {\bibinfo  {journal} {Nat. Phys.}\ }\textbf {\bibinfo {volume}
  {9}},\ \bibinfo {pages} {480} (\bibinfo {year} {2013})}\BibitemShut {NoStop}%
\bibitem [{\citenamefont {Mahboob}\ \emph {et~al.}(2012)\citenamefont
  {Mahboob}, \citenamefont {Nishiguchi}, \citenamefont {Okamoto},\ and\
  \citenamefont {Yamaguchi}}]{Mahboob2012}%
  \BibitemOpen
  \bibfield  {author} {\bibinfo {author} {\bibfnamefont {I.}~\bibnamefont
  {Mahboob}}, \bibinfo {author} {\bibfnamefont {K.}~\bibnamefont {Nishiguchi}},
  \bibinfo {author} {\bibfnamefont {H.}~\bibnamefont {Okamoto}}, \ and\
  \bibinfo {author} {\bibfnamefont {H.}~\bibnamefont {Yamaguchi}},\ }\bibfield
  {title} {\enquote {\bibinfo {title} {Phonon-cavity electromechanics},}\
  }\href {\doibase 10.1038/nphys2277} {\bibfield  {journal} {\bibinfo
  {journal} {Nat. Phys.}\ }\textbf {\bibinfo {volume} {8}},\ \bibinfo {pages}
  {387} (\bibinfo {year} {2012})}\BibitemShut {NoStop}%
\bibitem [{\citenamefont {Xu}\ \emph {et~al.}(2021)\citenamefont {Xu},
  \citenamefont {Lai}, \citenamefont {Qian}, \citenamefont {Hou}, \citenamefont
  {Miranowicz},\ and\ \citenamefont {Nori}}]{Xu-2021}%
  \BibitemOpen
  \bibfield  {author} {\bibinfo {author} {\bibfnamefont {Hai}\ \bibnamefont
  {Xu}}, \bibinfo {author} {\bibfnamefont {Deng-Gao}\ \bibnamefont {Lai}},
  \bibinfo {author} {\bibfnamefont {Yi-Bing}\ \bibnamefont {Qian}}, \bibinfo
  {author} {\bibfnamefont {Bang-Pin}\ \bibnamefont {Hou}}, \bibinfo {author}
  {\bibfnamefont {Adam}\ \bibnamefont {Miranowicz}}, \ and\ \bibinfo {author}
  {\bibfnamefont {Franco}\ \bibnamefont {Nori}},\ }\bibfield  {title} {\enquote
  {\bibinfo {title} {Optomechanical dynamics in the $\mathcal{PT}$- and
  broken-$\mathcal{PT}$-symmetric regimes},}\ }\href {\doibase
  10.1103/PhysRevA.104.053518} {\bibfield  {journal} {\bibinfo  {journal}
  {Phys. Rev. A}\ }\textbf {\bibinfo {volume} {104}},\ \bibinfo {pages}
  {053518} (\bibinfo {year} {2021})}\BibitemShut {NoStop}%
\bibitem [{\citenamefont {Goos}\ and\ \citenamefont
  {Hänchen}(1943)}]{Goos-1943}%
  \BibitemOpen
  \bibfield  {author} {\bibinfo {author} {\bibfnamefont {F.}~\bibnamefont
  {Goos}}\ and\ \bibinfo {author} {\bibfnamefont {H.}~\bibnamefont
  {Hänchen}},\ }\bibfield  {title} {\enquote {\bibinfo {title} {Über das
  eindringen des totalreflektierten lichtes in das dünnere medium},}\ }\href
  {\doibase https://doi.org/10.1002/andp.19434350504} {\bibfield  {journal}
  {\bibinfo  {journal} {Ann. Phys. (Leipzig)}\ }\textbf {\bibinfo {volume}
  {435}},\ \bibinfo {pages} {383} (\bibinfo {year} {1943})}\BibitemShut
  {NoStop}%
\bibitem [{\citenamefont {Goos}\ and\ \citenamefont
  {H\"{a}nchen}(1947)}]{Goos-1947}%
  \BibitemOpen
  \bibfield  {author} {\bibinfo {author} {\bibfnamefont {F.}~\bibnamefont
  {Goos}}\ and\ \bibinfo {author} {\bibfnamefont {H.}~\bibnamefont
  {H\"{a}nchen}},\ }\bibfield  {title} {\enquote {\bibinfo {title} {Ein neuer
  und fundamentaler versuch zur totalreflexion},}\ }\href {\doibase
  https://doi.org/10.1002/andp.19474360704} {\bibfield  {journal} {\bibinfo
  {journal} {Ann. Phys. (Leipzig)}\ }\textbf {\bibinfo {volume} {436}},\
  \bibinfo {pages} {333} (\bibinfo {year} {1947})}\BibitemShut {NoStop}%
\bibitem [{\citenamefont {Renard}(1964)}]{Renard1964}%
  \BibitemOpen
  \bibfield  {author} {\bibinfo {author} {\bibfnamefont {R\'{e}mi~H.}\
  \bibnamefont {Renard}},\ }\bibfield  {title} {\enquote {\bibinfo {title}
  {Total reflection: A new evaluation of the {Goos-H\"{a}nchen} shift},}\
  }\href {\doibase 10.1364/JOSA.54.001190} {\bibfield  {journal} {\bibinfo
  {journal} {J. Opt. Soc. Am.}\ }\textbf {\bibinfo {volume} {54}},\ \bibinfo
  {pages} {1190} (\bibinfo {year} {1964})}\BibitemShut {NoStop}%
\bibitem [{\citenamefont {Bliokh}\ and\ \citenamefont
  {Aiello}(2013)}]{Bliokh2013}%
  \BibitemOpen
  \bibfield  {author} {\bibinfo {author} {\bibfnamefont {K~Y}\ \bibnamefont
  {Bliokh}}\ and\ \bibinfo {author} {\bibfnamefont {A}~\bibnamefont {Aiello}},\
  }\bibfield  {title} {\enquote {\bibinfo {title} {{{Goos-H\"{a}nchen} and
  Imbert–Fedorov beam shifts: an overview}},}\ }\href {\doibase
  10.1088/2040-8978/15/1/014001} {\bibfield  {journal} {\bibinfo  {journal} {J.
  Opt.}\ }\textbf {\bibinfo {volume} {15}},\ \bibinfo {pages} {014001}
  (\bibinfo {year} {2013})}\BibitemShut {NoStop}%
\bibitem [{\citenamefont {Schomerus}\ and\ \citenamefont
  {Hentschel}(2006)}]{Schomerus2006}%
  \BibitemOpen
  \bibfield  {author} {\bibinfo {author} {\bibfnamefont {Henning}\ \bibnamefont
  {Schomerus}}\ and\ \bibinfo {author} {\bibfnamefont {Martina}\ \bibnamefont
  {Hentschel}},\ }\bibfield  {title} {\enquote {\bibinfo {title} {Correcting
  ray optics at curved dielectric microresonator interfaces: Phase-space
  unification of {Fresnel} filtering and the {Goos-H\"anchen} shift},}\ }\href
  {\doibase 10.1103/PhysRevLett.96.243903} {\bibfield  {journal} {\bibinfo
  {journal} {Phys. Rev. Lett.}\ }\textbf {\bibinfo {volume} {96}},\ \bibinfo
  {pages} {243903} (\bibinfo {year} {2006})}\BibitemShut {NoStop}%
\bibitem [{\citenamefont {de~Fornel}(2001)}]{deFornel2001}%
  \BibitemOpen
  \bibfield  {author} {\bibinfo {author} {\bibfnamefont {Fr{\'e}d{\'e}rique}\
  \bibnamefont {de~Fornel}},\ }\href@noop {} {\emph {\bibinfo {title}
  {{Evanescent waves : from Newtonian optics to atomic optics}}}}\ (\bibinfo
  {publisher} {Springer},\ \bibinfo {address} {Berlin},\ \bibinfo {year}
  {2001})\BibitemShut {NoStop}%
\bibitem [{\citenamefont {Yin}\ and\ \citenamefont
  {Hesselink}(2006)}]{Yin2006}%
  \BibitemOpen
  \bibfield  {author} {\bibinfo {author} {\bibfnamefont {Xiaobo}\ \bibnamefont
  {Yin}}\ and\ \bibinfo {author} {\bibfnamefont {Lambertus}\ \bibnamefont
  {Hesselink}},\ }\bibfield  {title} {\enquote {\bibinfo {title}
  {{Goos-H\"{a}nchen} shift surface plasmon resonance sensor},}\ }\href
  {\doibase 10.1063/1.2424277} {\bibfield  {journal} {\bibinfo  {journal}
  {Appl. Phys. Lett.}\ }\textbf {\bibinfo {volume} {89}},\ \bibinfo {pages}
  {261108} (\bibinfo {year} {2006})}\BibitemShut {NoStop}%
\bibitem [{\citenamefont {Soboleva}\ \emph {et~al.}(2012)\citenamefont
  {Soboleva}, \citenamefont {Moskalenko},\ and\ \citenamefont
  {Fedyanin}}]{Soboleva-2012}%
  \BibitemOpen
  \bibfield  {author} {\bibinfo {author} {\bibfnamefont {I.~V.}\ \bibnamefont
  {Soboleva}}, \bibinfo {author} {\bibfnamefont {V.~V.}\ \bibnamefont
  {Moskalenko}}, \ and\ \bibinfo {author} {\bibfnamefont {A.~A.}\ \bibnamefont
  {Fedyanin}},\ }\bibfield  {title} {\enquote {\bibinfo {title} {Giant
  {Goos-H\"anchen} effect and fano resonance at photonic crystal surfaces},}\
  }\href {\doibase 10.1103/PhysRevLett.108.123901} {\bibfield  {journal}
  {\bibinfo  {journal} {Phys. Rev. Lett.}\ }\textbf {\bibinfo {volume} {108}},\
  \bibinfo {pages} {123901} (\bibinfo {year} {2012})}\BibitemShut {NoStop}%
\bibitem [{\citenamefont {de~Haan}\ \emph {et~al.}(2010)\citenamefont
  {de~Haan}, \citenamefont {Plomp}, \citenamefont {Rekveldt}, \citenamefont
  {Kraan}, \citenamefont {van Well}, \citenamefont {Dalgliesh},\ and\
  \citenamefont {Langridge}}]{Hann-PRL}%
  \BibitemOpen
  \bibfield  {author} {\bibinfo {author} {\bibfnamefont {Victor-O.}\
  \bibnamefont {de~Haan}}, \bibinfo {author} {\bibfnamefont {Jeroen}\
  \bibnamefont {Plomp}}, \bibinfo {author} {\bibfnamefont {Theo~M.}\
  \bibnamefont {Rekveldt}}, \bibinfo {author} {\bibfnamefont {Wicher~H.}\
  \bibnamefont {Kraan}}, \bibinfo {author} {\bibfnamefont {Ad~A.}\ \bibnamefont
  {van Well}}, \bibinfo {author} {\bibfnamefont {Robert~M.}\ \bibnamefont
  {Dalgliesh}}, \ and\ \bibinfo {author} {\bibfnamefont {Sean}\ \bibnamefont
  {Langridge}},\ }\bibfield  {title} {\enquote {\bibinfo {title} {Observation
  of the {Goos-H\"{a}nchen} shift with neutrons},}\ }\href {\doibase
  10.1103/PhysRevLett.104.010401} {\bibfield  {journal} {\bibinfo  {journal}
  {Phys. Rev. Lett.}\ }\textbf {\bibinfo {volume} {104}},\ \bibinfo {pages}
  {010401} (\bibinfo {year} {2010})}\BibitemShut {NoStop}%
\bibitem [{\citenamefont {Beenakker}\ \emph {et~al.}(2009)\citenamefont
  {Beenakker}, \citenamefont {Sepkhanov}, \citenamefont {Akhmerov},\ and\
  \citenamefont {Tworzyd\l{}o}}]{Beenakker-PRL}%
  \BibitemOpen
  \bibfield  {author} {\bibinfo {author} {\bibfnamefont {C.~W.~J.}\
  \bibnamefont {Beenakker}}, \bibinfo {author} {\bibfnamefont {R.~A.}\
  \bibnamefont {Sepkhanov}}, \bibinfo {author} {\bibfnamefont {A.~R.}\
  \bibnamefont {Akhmerov}}, \ and\ \bibinfo {author} {\bibfnamefont
  {J.}~\bibnamefont {Tworzyd\l{}o}},\ }\bibfield  {title} {\enquote {\bibinfo
  {title} {Quantum {Goos-H\"{a}nchen} effect in graphene},}\ }\href {\doibase
  10.1103/PhysRevLett.102.146804} {\bibfield  {journal} {\bibinfo  {journal}
  {Phys. Rev. Lett.}\ }\textbf {\bibinfo {volume} {102}},\ \bibinfo {pages}
  {146804} (\bibinfo {year} {2009})}\BibitemShut {NoStop}%
\bibitem [{\citenamefont {Wu}\ \emph {et~al.}(2011)\citenamefont {Wu},
  \citenamefont {Zhai}, \citenamefont {Peeters}, \citenamefont {Xu},\ and\
  \citenamefont {Chang}}]{Wu-PRL}%
  \BibitemOpen
  \bibfield  {author} {\bibinfo {author} {\bibfnamefont {Zhenhua}\ \bibnamefont
  {Wu}}, \bibinfo {author} {\bibfnamefont {F.}~\bibnamefont {Zhai}}, \bibinfo
  {author} {\bibfnamefont {F.~M.}\ \bibnamefont {Peeters}}, \bibinfo {author}
  {\bibfnamefont {H.~Q.}\ \bibnamefont {Xu}}, \ and\ \bibinfo {author}
  {\bibfnamefont {Kai}\ \bibnamefont {Chang}},\ }\bibfield  {title} {\enquote
  {\bibinfo {title} {Valley-dependent brewster angles and goos-h\"{a}nchen
  effect in strained graphene},}\ }\href {\doibase
  10.1103/PhysRevLett.106.176802} {\bibfield  {journal} {\bibinfo  {journal}
  {Phys. Rev. Lett.}\ }\textbf {\bibinfo {volume} {106}},\ \bibinfo {pages}
  {176802} (\bibinfo {year} {2011})}\BibitemShut {NoStop}%
\bibitem [{\citenamefont {Dadoenkova}\ \emph {et~al.}(2012)\citenamefont
  {Dadoenkova}, \citenamefont {Dadoenkova}, \citenamefont {Lyubchanskii},
  \citenamefont {Sokolovskyy}, \citenamefont {Kłos}, \citenamefont
  {Romero-Vivas},\ and\ \citenamefont {Krawczyk}}]{Dadoenkova-2012}%
  \BibitemOpen
  \bibfield  {author} {\bibinfo {author} {\bibfnamefont {Yu.~S.}\ \bibnamefont
  {Dadoenkova}}, \bibinfo {author} {\bibfnamefont {N.~N.}\ \bibnamefont
  {Dadoenkova}}, \bibinfo {author} {\bibfnamefont {I.~L.}\ \bibnamefont
  {Lyubchanskii}}, \bibinfo {author} {\bibfnamefont {M.~L.}\ \bibnamefont
  {Sokolovskyy}}, \bibinfo {author} {\bibfnamefont {J.~W.}\ \bibnamefont
  {Kłos}}, \bibinfo {author} {\bibfnamefont {J.}~\bibnamefont {Romero-Vivas}},
  \ and\ \bibinfo {author} {\bibfnamefont {M.}~\bibnamefont {Krawczyk}},\
  }\bibfield  {title} {\enquote {\bibinfo {title} {Huge {Goos-H\"{a}nchen}
  effect for spin waves: A promising tool for study magnetic properties at
  interfaces},}\ }\href {\doibase 10.1063/1.4738987} {\bibfield  {journal}
  {\bibinfo  {journal} {Appl. Phys. Lett.}\ }\textbf {\bibinfo {volume}
  {101}},\ \bibinfo {pages} {042404} (\bibinfo {year} {2012})}\BibitemShut
  {NoStop}%
\bibitem [{\citenamefont {Jiang}\ \emph {et~al.}(2015)\citenamefont {Jiang},
  \citenamefont {Jiang}, \citenamefont {Liu}, \citenamefont {Sun},\ and\
  \citenamefont {Xie}}]{Jiang-PRL}%
  \BibitemOpen
  \bibfield  {author} {\bibinfo {author} {\bibfnamefont {Qing-Dong}\
  \bibnamefont {Jiang}}, \bibinfo {author} {\bibfnamefont {Hua}\ \bibnamefont
  {Jiang}}, \bibinfo {author} {\bibfnamefont {Haiwen}\ \bibnamefont {Liu}},
  \bibinfo {author} {\bibfnamefont {Qing-Feng}\ \bibnamefont {Sun}}, \ and\
  \bibinfo {author} {\bibfnamefont {X.~C.}\ \bibnamefont {Xie}},\ }\bibfield
  {title} {\enquote {\bibinfo {title} {Topological {Imbert-Fedorov} shift in
  weyl semimetals},}\ }\href {\doibase 10.1103/PhysRevLett.115.156602}
  {\bibfield  {journal} {\bibinfo  {journal} {Phys. Rev. Lett.}\ }\textbf
  {\bibinfo {volume} {115}},\ \bibinfo {pages} {156602} (\bibinfo {year}
  {2015})}\BibitemShut {NoStop}%
\bibitem [{\citenamefont {Chattopadhyay}\ \emph {et~al.}(2019)\citenamefont
  {Chattopadhyay}, \citenamefont {Shi}, \citenamefont {Zhang}, \citenamefont
  {Song},\ and\ \citenamefont {Chong}}]{Chattopadhyay2019}%
  \BibitemOpen
  \bibfield  {author} {\bibinfo {author} {\bibfnamefont {Udvas}\ \bibnamefont
  {Chattopadhyay}}, \bibinfo {author} {\bibfnamefont {Li-kun}\ \bibnamefont
  {Shi}}, \bibinfo {author} {\bibfnamefont {Baile}\ \bibnamefont {Zhang}},
  \bibinfo {author} {\bibfnamefont {Justin C.~W.}\ \bibnamefont {Song}}, \ and\
  \bibinfo {author} {\bibfnamefont {Y.~D.}\ \bibnamefont {Chong}},\ }\bibfield
  {title} {\enquote {\bibinfo {title} {{Fermi-Arc-Induced Vortex Structure in
  Weyl Beam Shifts}},}\ }\href {\doibase 10.1103/PhysRevLett.122.066602}
  {\bibfield  {journal} {\bibinfo  {journal} {Phys. Rev. Lett.}\ }\textbf
  {\bibinfo {volume} {122}},\ \bibinfo {pages} {066602} (\bibinfo {year}
  {2019})}\BibitemShut {NoStop}%
\bibitem [{\citenamefont {Liu}\ \emph {et~al.}(2020)\citenamefont {Liu},
  \citenamefont {Yu}, \citenamefont {Xiao},\ and\ \citenamefont
  {Yang}}]{Liu-2020}%
  \BibitemOpen
  \bibfield  {author} {\bibinfo {author} {\bibfnamefont {Ying}\ \bibnamefont
  {Liu}}, \bibinfo {author} {\bibfnamefont {Zhi-Ming}\ \bibnamefont {Yu}},
  \bibinfo {author} {\bibfnamefont {Cong}\ \bibnamefont {Xiao}}, \ and\
  \bibinfo {author} {\bibfnamefont {Shengyuan~A.}\ \bibnamefont {Yang}},\
  }\bibfield  {title} {\enquote {\bibinfo {title} {Quantized circulation of
  anomalous shift in interface reflection},}\ }\href {\doibase
  10.1103/PhysRevLett.125.076801} {\bibfield  {journal} {\bibinfo  {journal}
  {Phys. Rev. Lett.}\ }\textbf {\bibinfo {volume} {125}},\ \bibinfo {pages}
  {076801} (\bibinfo {year} {2020})}\BibitemShut {NoStop}%
\bibitem [{\citenamefont {Huang}\ \emph {et~al.}(2008)\citenamefont {Huang},
  \citenamefont {Duan}, \citenamefont {Ling},\ and\ \citenamefont
  {Zhang}}]{Huang-PRL}%
  \BibitemOpen
  \bibfield  {author} {\bibinfo {author} {\bibfnamefont {Jianhua}\ \bibnamefont
  {Huang}}, \bibinfo {author} {\bibfnamefont {Zhenglu}\ \bibnamefont {Duan}},
  \bibinfo {author} {\bibfnamefont {Hong~Y.}\ \bibnamefont {Ling}}, \ and\
  \bibinfo {author} {\bibfnamefont {Weiping}\ \bibnamefont {Zhang}},\
  }\bibfield  {title} {\enquote {\bibinfo {title} {{Goos-H\"{a}nchen}-like
  shifts in atom optics},}\ }\href {\doibase 10.1103/PhysRevA.77.063608}
  {\bibfield  {journal} {\bibinfo  {journal} {Phys. Rev. A}\ }\textbf {\bibinfo
  {volume} {77}},\ \bibinfo {pages} {063608} (\bibinfo {year}
  {2008})}\BibitemShut {NoStop}%
\bibitem [{\citenamefont {Lee}\ \emph {et~al.}(2014)\citenamefont {Lee},
  \citenamefont {Le~Deunff}, \citenamefont {Choi},\ and\ \citenamefont
  {Ketzmerick}}]{Lee-2014}%
  \BibitemOpen
  \bibfield  {author} {\bibinfo {author} {\bibfnamefont {Soo-Young}\
  \bibnamefont {Lee}}, \bibinfo {author} {\bibfnamefont {J\'er\'emy}\
  \bibnamefont {Le~Deunff}}, \bibinfo {author} {\bibfnamefont {Muhan}\
  \bibnamefont {Choi}}, \ and\ \bibinfo {author} {\bibfnamefont {Roland}\
  \bibnamefont {Ketzmerick}},\ }\bibfield  {title} {\enquote {\bibinfo {title}
  {Quantum {Goos-H\"{a}nchen} shift and tunneling transmission at a curved step
  potential},}\ }\href {\doibase 10.1103/PhysRevA.89.022120} {\bibfield
  {journal} {\bibinfo  {journal} {Phys. Rev. A}\ }\textbf {\bibinfo {volume}
  {89}},\ \bibinfo {pages} {022120} (\bibinfo {year} {2014})}\BibitemShut
  {NoStop}%
\bibitem [{\citenamefont {Han}\ \emph {et~al.}(2021)\citenamefont {Han},
  \citenamefont {Li}, \citenamefont {Zhou}, \citenamefont {Jiang},
  \citenamefont {Chang}, \citenamefont {Huang}, \citenamefont {Zhang},\ and\
  \citenamefont {Xiao}}]{Han2021}%
  \BibitemOpen
  \bibfield  {author} {\bibinfo {author} {\bibfnamefont {Peng}\ \bibnamefont
  {Han}}, \bibinfo {author} {\bibfnamefont {Wenxiu}\ \bibnamefont {Li}},
  \bibinfo {author} {\bibfnamefont {Yang}\ \bibnamefont {Zhou}}, \bibinfo
  {author} {\bibfnamefont {Shuo}\ \bibnamefont {Jiang}}, \bibinfo {author}
  {\bibfnamefont {Xiaoyang}\ \bibnamefont {Chang}}, \bibinfo {author}
  {\bibfnamefont {Anping}\ \bibnamefont {Huang}}, \bibinfo {author}
  {\bibfnamefont {Hao}\ \bibnamefont {Zhang}}, \ and\ \bibinfo {author}
  {\bibfnamefont {Zhisong}\ \bibnamefont {Xiao}},\ }\bibfield  {title}
  {\enquote {\bibinfo {title} {Giant and tunable {Goos-H{\"a}nchen} shift with
  a high reflectance induced by $\mathcal{PT}$-symmetry in atomic vapor},}\
  }\href {\doibase 10.1364/OE.432082} {\bibfield  {journal} {\bibinfo
  {journal} {Opt. Express.}\ }\textbf {\bibinfo {volume} {29}},\ \bibinfo
  {pages} {30436} (\bibinfo {year} {2021})}\BibitemShut {NoStop}%
\bibitem [{\citenamefont {Yue}\ \emph {et~al.}(2021)\citenamefont {Yue},
  \citenamefont {Zhen}, \citenamefont {Ding}, \citenamefont {Zhou},\ and\
  \citenamefont {Deng}}]{Yue2021}%
  \BibitemOpen
  \bibfield  {author} {\bibinfo {author} {\bibfnamefont {Qinxin}\ \bibnamefont
  {Yue}}, \bibinfo {author} {\bibfnamefont {Weiming}\ \bibnamefont {Zhen}},
  \bibinfo {author} {\bibfnamefont {Yiping}\ \bibnamefont {Ding}}, \bibinfo
  {author} {\bibfnamefont {Xiang}\ \bibnamefont {Zhou}}, \ and\ \bibinfo
  {author} {\bibfnamefont {Dongmei}\ \bibnamefont {Deng}},\ }\bibfield  {title}
  {\enquote {\bibinfo {title} {Giant {Goos-H{\"a}nchen} shifts controlled by
  exceptional points in a $\mathcal{PT}$-symmetric periodic multilayered
  structure coated with graphene},}\ }\href {\doibase 10.1364/OME.441184}
  {\bibfield  {journal} {\bibinfo  {journal} {Opt. Mater. Express}\ }\textbf
  {\bibinfo {volume} {11}},\ \bibinfo {pages} {3954} (\bibinfo {year}
  {2021})}\BibitemShut {NoStop}%
\bibitem [{\citenamefont {Ma}\ and\ \citenamefont {Gao}(2017)}]{Ma2017}%
  \BibitemOpen
  \bibfield  {author} {\bibinfo {author} {\bibfnamefont {Pujuan}\ \bibnamefont
  {Ma}}\ and\ \bibinfo {author} {\bibfnamefont {Lei}\ \bibnamefont {Gao}},\
  }\bibfield  {title} {\enquote {\bibinfo {title} {Large and tunable lateral
  shifts in one-dimensional $\mathcal{PT}$-symmetric layered structures},}\
  }\href {\doibase 10.1364/OE.25.009676} {\bibfield  {journal} {\bibinfo
  {journal} {Opt. Express}\ }\textbf {\bibinfo {volume} {25}},\ \bibinfo
  {pages} {9676} (\bibinfo {year} {2017})}\BibitemShut {NoStop}%
\bibitem [{\citenamefont {{Zhang}}\ \emph {et~al.}(2022)\citenamefont
  {{Zhang}}, \citenamefont {{Zeng}}, \citenamefont {{Zhou}},\ and\
  \citenamefont {{Ni}}}]{Zhang2022}%
  \BibitemOpen
  \bibfield  {author} {\bibinfo {author} {\bibfnamefont {Xiaoping}\
  \bibnamefont {{Zhang}}}, \bibinfo {author} {\bibfnamefont {Xiangjin}\
  \bibnamefont {{Zeng}}}, \bibinfo {author} {\bibfnamefont {Guopeng}\
  \bibnamefont {{Zhou}}}, \ and\ \bibinfo {author} {\bibfnamefont {Hao}\
  \bibnamefont {{Ni}}},\ }\bibfield  {title} {\enquote {\bibinfo {title}
  {{Tunable spatial {Goos-H{\"a}nchen} shift in periodic
  $\mathcal{PT}$-symmetric photonic crystals with a central defect}},}\ }\href
  {\doibase 10.1088/1402-4896/ac9ca1} {\bibfield  {journal} {\bibinfo
  {journal} {Phys. Scr.}\ }\textbf {\bibinfo {volume} {97}},\ \bibinfo {eid}
  {125503} (\bibinfo {year} {2022})}\BibitemShut {NoStop}%
\bibitem [{\citenamefont {Longhi}\ \emph {et~al.}(2011)\citenamefont {Longhi},
  \citenamefont {Della~Valle},\ and\ \citenamefont {Staliunas}}]{Longhi2011}%
  \BibitemOpen
  \bibfield  {author} {\bibinfo {author} {\bibfnamefont {Stefano}\ \bibnamefont
  {Longhi}}, \bibinfo {author} {\bibfnamefont {Giuseppe}\ \bibnamefont
  {Della~Valle}}, \ and\ \bibinfo {author} {\bibfnamefont {Kestutis}\
  \bibnamefont {Staliunas}},\ }\bibfield  {title} {\enquote {\bibinfo {title}
  {{Goos-H\"{a}nchen} shift in complex crystals},}\ }\href {\doibase
  10.1103/PhysRevA.84.042119} {\bibfield  {journal} {\bibinfo  {journal} {Phys.
  Rev. A}\ }\textbf {\bibinfo {volume} {84}},\ \bibinfo {pages} {042119}
  (\bibinfo {year} {2011})}\BibitemShut {NoStop}%
\bibitem [{\citenamefont {Ziauddin}\ \emph {et~al.}(2015)\citenamefont
  {Ziauddin}, \citenamefont {Chuang},\ and\ \citenamefont
  {Lee}}]{Ziauddin2015}%
  \BibitemOpen
  \bibfield  {author} {\bibinfo {author} {\bibnamefont {Ziauddin}}, \bibinfo
  {author} {\bibfnamefont {You-Lin}\ \bibnamefont {Chuang}}, \ and\ \bibinfo
  {author} {\bibfnamefont {Ray-Kuang}\ \bibnamefont {Lee}},\ }\bibfield
  {title} {\enquote {\bibinfo {title} {Giant {Goos-H\"anchen} shift using
  $\mathcal{PT}$ symmetry},}\ }\href {\doibase 10.1103/PhysRevA.92.013815}
  {\bibfield  {journal} {\bibinfo  {journal} {Phys. Rev. A}\ }\textbf {\bibinfo
  {volume} {92}},\ \bibinfo {pages} {013815} (\bibinfo {year}
  {2015})}\BibitemShut {NoStop}%
\bibitem [{\citenamefont {Fahad}\ and\ \citenamefont
  {Xianlong}(2025)}]{fahad2025}%
  \BibitemOpen
  \bibfield  {author} {\bibinfo {author} {\bibfnamefont {Shah}\ \bibnamefont
  {Fahad}}\ and\ \bibinfo {author} {\bibfnamefont {Gao}\ \bibnamefont
  {Xianlong}},\ }\href {https://arxiv.org/abs/2511.20262} {\enquote {\bibinfo
  {title} {$\mathcal{PT}$-assisted control of {Goos-H{\"a}nchen} shift in
  cavity magnomechanics},}\ } (\bibinfo {year} {2025}),\ \Eprint
  {http://arxiv.org/abs/2511.20262} {arXiv:2511.20262} \BibitemShut {NoStop}%
\bibitem [{\citenamefont {Ullah}\ \emph {et~al.}(2019)\citenamefont {Ullah},
  \citenamefont {Abbas}, \citenamefont {Jing},\ and\ \citenamefont
  {Wang}}]{Muhib-2019}%
  \BibitemOpen
  \bibfield  {author} {\bibinfo {author} {\bibfnamefont {Muhib}\ \bibnamefont
  {Ullah}}, \bibinfo {author} {\bibfnamefont {Adeel}\ \bibnamefont {Abbas}},
  \bibinfo {author} {\bibfnamefont {Jun}\ \bibnamefont {Jing}}, \ and\ \bibinfo
  {author} {\bibfnamefont {Li-Gang}\ \bibnamefont {Wang}},\ }\bibfield  {title}
  {\enquote {\bibinfo {title} {Flexible manipulation of the {Goos-H\"{a}nchen}
  shift in a cavity optomechanical system},}\ }\href {\doibase
  10.1103/PhysRevA.100.063833} {\bibfield  {journal} {\bibinfo  {journal}
  {Phys. Rev. A}\ }\textbf {\bibinfo {volume} {100}},\ \bibinfo {pages}
  {063833} (\bibinfo {year} {2019})}\BibitemShut {NoStop}%
\bibitem [{\citenamefont {Ghaisuddin}\ \emph {et~al.}(2021)\citenamefont
  {Ghaisuddin}, \citenamefont {Abbas}, \citenamefont {Ali~Khan}, \citenamefont
  {Ali},\ and\ \citenamefont {Ziauddin}}]{Ghaisuddin_2021}%
  \BibitemOpen
  \bibfield  {author} {\bibinfo {author} {\bibnamefont {Ghaisuddin}}, \bibinfo
  {author} {\bibfnamefont {Muqaddar}\ \bibnamefont {Abbas}}, \bibinfo {author}
  {\bibfnamefont {Anwar}\ \bibnamefont {Ali~Khan}}, \bibinfo {author}
  {\bibfnamefont {Hazrat}\ \bibnamefont {Ali}}, \ and\ \bibinfo {author}
  {\bibnamefont {Ziauddin}},\ }\bibfield  {title} {\enquote {\bibinfo {title}
  {Enhancement of the {Goos-H\"{a}nchen} shift in an optomechancal cavity via
  casimir force},}\ }\href {\doibase 10.1088/1402-4896/ac1dca} {\bibfield
  {journal} {\bibinfo  {journal} {Phys. Scr.}\ }\textbf {\bibinfo {volume}
  {96}},\ \bibinfo {pages} {125104} (\bibinfo {year} {2021})}\BibitemShut
  {NoStop}%
\bibitem [{\citenamefont {Khan}\ \emph {et~al.}(2020)\citenamefont {Khan},
  \citenamefont {Abbas}, \citenamefont {Chaung}, \citenamefont {Ahmed},\ and\
  \citenamefont {Ziauddin}}]{Anwar-2020}%
  \BibitemOpen
  \bibfield  {author} {\bibinfo {author} {\bibfnamefont {Anwar~Ali}\
  \bibnamefont {Khan}}, \bibinfo {author} {\bibfnamefont {Muqaddar}\
  \bibnamefont {Abbas}}, \bibinfo {author} {\bibfnamefont {You-Lin}\
  \bibnamefont {Chaung}}, \bibinfo {author} {\bibfnamefont {Iftikhar}\
  \bibnamefont {Ahmed}}, \ and\ \bibinfo {author} {\bibnamefont {Ziauddin}},\
  }\bibfield  {title} {\enquote {\bibinfo {title} {Investigation of the
  {Goos-H\"{a}nchen} shift in an optomechanical cavity via quantum control},}\
  }\href {\doibase 10.1103/PhysRevA.102.053718} {\bibfield  {journal} {\bibinfo
   {journal} {Phys. Rev. A}\ }\textbf {\bibinfo {volume} {102}},\ \bibinfo
  {pages} {053718} (\bibinfo {year} {2020})}\BibitemShut {NoStop}%
\bibitem [{\citenamefont {Cai}\ \emph {et~al.}(2000)\citenamefont {Cai},
  \citenamefont {Painter},\ and\ \citenamefont {Vahala}}]{Cai-PRL}%
  \BibitemOpen
  \bibfield  {author} {\bibinfo {author} {\bibfnamefont {Ming}\ \bibnamefont
  {Cai}}, \bibinfo {author} {\bibfnamefont {Oskar}\ \bibnamefont {Painter}}, \
  and\ \bibinfo {author} {\bibfnamefont {Kerry~J.}\ \bibnamefont {Vahala}},\
  }\bibfield  {title} {\enquote {\bibinfo {title} {{Observation of Critical
  Coupling in a Fiber Taper to a Silica-Microsphere Whispering-Gallery Mode
  System}},}\ }\href {\doibase 10.1103/PhysRevLett.85.74} {\bibfield  {journal}
  {\bibinfo  {journal} {Phys. Rev. Lett.}\ }\textbf {\bibinfo {volume} {85}},\
  \bibinfo {pages} {74} (\bibinfo {year} {2000})}\BibitemShut {NoStop}%
\bibitem [{\citenamefont {Spillane}\ \emph {et~al.}(2003)\citenamefont
  {Spillane}, \citenamefont {Kippenberg}, \citenamefont {Painter},\ and\
  \citenamefont {Vahala}}]{Spillane-PRL}%
  \BibitemOpen
  \bibfield  {author} {\bibinfo {author} {\bibfnamefont {S.~M.}\ \bibnamefont
  {Spillane}}, \bibinfo {author} {\bibfnamefont {T.~J.}\ \bibnamefont
  {Kippenberg}}, \bibinfo {author} {\bibfnamefont {O.~J.}\ \bibnamefont
  {Painter}}, \ and\ \bibinfo {author} {\bibfnamefont {K.~J.}\ \bibnamefont
  {Vahala}},\ }\bibfield  {title} {\enquote {\bibinfo {title} {{Ideality in a
  Fiber-Taper-Coupled Microresonator System for Application to Cavity Quantum
  Electrodynamics}},}\ }\href {\doibase 10.1103/PhysRevLett.91.043902}
  {\bibfield  {journal} {\bibinfo  {journal} {Phys. Rev. Lett.}\ }\textbf
  {\bibinfo {volume} {91}},\ \bibinfo {pages} {043902} (\bibinfo {year}
  {2003})}\BibitemShut {NoStop}%
\bibitem [{\citenamefont {Xiong}\ \emph {et~al.}(2015)\citenamefont {Xiong},
  \citenamefont {Si}, \citenamefont {Lv}, \citenamefont {Yang},\ and\
  \citenamefont {Wu}}]{Xiong2015}%
  \BibitemOpen
  \bibfield  {author} {\bibinfo {author} {\bibfnamefont {Hao}\ \bibnamefont
  {Xiong}}, \bibinfo {author} {\bibfnamefont {LiuGang}\ \bibnamefont {Si}},
  \bibinfo {author} {\bibfnamefont {XinYou}\ \bibnamefont {Lv}}, \bibinfo
  {author} {\bibfnamefont {XiaoXue}\ \bibnamefont {Yang}}, \ and\ \bibinfo
  {author} {\bibfnamefont {Ying}\ \bibnamefont {Wu}},\ }\bibfield  {title}
  {\enquote {\bibinfo {title} {{Review of cavity optomechanics in the
  weak-coupling regime: from linearization to intrinsic nonlinear
  interactions}},}\ }\href {\doibase 10.1007/s11433-015-5648-9} {\bibfield
  {journal} {\bibinfo  {journal} {Sci China Phys Mech Astron.}\ }\textbf
  {\bibinfo {volume} {58}},\ \bibinfo {pages} {1} (\bibinfo {year}
  {2015})}\BibitemShut {NoStop}%
\bibitem [{\citenamefont {{Li}}\ \emph {et~al.}(2018)\citenamefont {{Li}},
  \citenamefont {{Yang}}, \citenamefont {{Zhang}}, \citenamefont {{Shui}},
  \citenamefont {{Chen}},\ and\ \citenamefont {{Jiang}}}]{Ling2018}%
  \BibitemOpen
  \bibfield  {author} {\bibinfo {author} {\bibfnamefont {Ling}\ \bibnamefont
  {{Li}}}, \bibinfo {author} {\bibfnamefont {Wen-Xing}\ \bibnamefont {{Yang}}},
  \bibinfo {author} {\bibfnamefont {Yuexin}\ \bibnamefont {{Zhang}}}, \bibinfo
  {author} {\bibfnamefont {Tao}\ \bibnamefont {{Shui}}}, \bibinfo {author}
  {\bibfnamefont {Ai-Xi}\ \bibnamefont {{Chen}}}, \ and\ \bibinfo {author}
  {\bibfnamefont {Zhongming}\ \bibnamefont {{Jiang}}},\ }\bibfield  {title}
  {\enquote {\bibinfo {title} {{Enhanced generation of charge-dependent
  second-order sideband and high-sensitivity charge sensors in a
  gain-cavity-assisted optomechanical system}},}\ }\href {\doibase
  10.1103/PhysRevA.98.063840} {\bibfield  {journal} {\bibinfo  {journal} {Phys.
  Rev. A}\ }\textbf {\bibinfo {volume} {98}},\ \bibinfo {eid} {063840}
  (\bibinfo {year} {2018})}\BibitemShut {NoStop}%
\bibitem [{\citenamefont {Yin}\ \emph {et~al.}(2025)\citenamefont {Yin},
  \citenamefont {Wang}, \citenamefont {Peng}, \citenamefont {Zhang},
  \citenamefont {Wang}, \citenamefont {Lu}, \citenamefont {Wei},\ and\
  \citenamefont {Jing}}]{Yin2025}%
  \BibitemOpen
  \bibfield  {author} {\bibinfo {author} {\bibfnamefont {Bin}\ \bibnamefont
  {Yin}}, \bibinfo {author} {\bibfnamefont {Jie}\ \bibnamefont {Wang}},
  \bibinfo {author} {\bibfnamefont {Mei-Yu}\ \bibnamefont {Peng}}, \bibinfo
  {author} {\bibfnamefont {Qian}\ \bibnamefont {Zhang}}, \bibinfo {author}
  {\bibfnamefont {Deng}\ \bibnamefont {Wang}}, \bibinfo {author} {\bibfnamefont
  {Tian-Xiang}\ \bibnamefont {Lu}}, \bibinfo {author} {\bibfnamefont
  {Ke}~\bibnamefont {Wei}}, \ and\ \bibinfo {author} {\bibfnamefont {Hui}\
  \bibnamefont {Jing}},\ }\bibfield  {title} {\enquote {\bibinfo {title}
  {Molecular optomechanically induced transparency},}\ }\href {\doibase
  10.1103/PhysRevA.111.043507} {\bibfield  {journal} {\bibinfo  {journal}
  {Phys. Rev. A}\ }\textbf {\bibinfo {volume} {111}},\ \bibinfo {pages}
  {043507} (\bibinfo {year} {2025})}\BibitemShut {NoStop}%
\bibitem [{\citenamefont {Walls}\ and\ \citenamefont
  {Milburn}(1994)}]{walls1994quantum}%
  \BibitemOpen
  \bibfield  {author} {\bibinfo {author} {\bibfnamefont {D.~F.}\ \bibnamefont
  {Walls}}\ and\ \bibinfo {author} {\bibfnamefont {G.~J.}\ \bibnamefont
  {Milburn}},\ }\href {\doibase 10.1007/978-3-642-79504-6} {\emph {\bibinfo
  {title} {{Quantum Optics}}}}\ (\bibinfo  {publisher} {Springer-Verlag},\
  \bibinfo {address} {Berlin, Heidelberg},\ \bibinfo {year} {1994})\BibitemShut
  {NoStop}%
\bibitem [{\citenamefont {Artmann}(1948)}]{Artmann-1948}%
  \BibitemOpen
  \bibfield  {author} {\bibinfo {author} {\bibfnamefont {Kurt}\ \bibnamefont
  {Artmann}},\ }\bibfield  {title} {\enquote {\bibinfo {title} {Berechnung der
  seitenversetzung des totalreflektierten strahles},}\ }\href {\doibase
  https://doi.org/10.1002/andp.19484370108} {\bibfield  {journal} {\bibinfo
  {journal} {Ann. Phys. (Leipzig)}\ }\textbf {\bibinfo {volume} {437}},\
  \bibinfo {pages} {87} (\bibinfo {year} {1948})}\BibitemShut {NoStop}%
\bibitem [{\citenamefont {Li}(2003)}]{Li-2003PRL}%
  \BibitemOpen
  \bibfield  {author} {\bibinfo {author} {\bibfnamefont {Chun-Fang}\
  \bibnamefont {Li}},\ }\bibfield  {title} {\enquote {\bibinfo {title}
  {Negative lateral shift of a light beam transmitted through a dielectric slab
  and interaction of boundary effects},}\ }\href {\doibase
  10.1103/PhysRevLett.91.133903} {\bibfield  {journal} {\bibinfo  {journal}
  {Phys. Rev. Lett.}\ }\textbf {\bibinfo {volume} {91}},\ \bibinfo {pages}
  {133903} (\bibinfo {year} {2003})}\BibitemShut {NoStop}%
\bibitem [{\citenamefont {Wang}\ \emph {et~al.}(2005)\citenamefont {Wang},
  \citenamefont {Chen},\ and\ \citenamefont {Zhu}}]{Wang2005}%
  \BibitemOpen
  \bibfield  {author} {\bibinfo {author} {\bibfnamefont {Li-Gang}\ \bibnamefont
  {Wang}}, \bibinfo {author} {\bibfnamefont {Hong}\ \bibnamefont {Chen}}, \
  and\ \bibinfo {author} {\bibfnamefont {Shi-Yao}\ \bibnamefont {Zhu}},\
  }\bibfield  {title} {\enquote {\bibinfo {title} {Large negative{
  Goos-H\"{a}nchen} shift from a weakly absorbing dielectric slab},}\ }\href
  {\doibase 10.1364/OL.30.002936} {\bibfield  {journal} {\bibinfo  {journal}
  {Opt. Lett.}\ }\textbf {\bibinfo {volume} {30}},\ \bibinfo {pages} {2936}
  (\bibinfo {year} {2005})}\BibitemShut {NoStop}%
\bibitem [{\citenamefont {Wang}\ \emph {et~al.}(2008)\citenamefont {Wang},
  \citenamefont {Ikram},\ and\ \citenamefont {Zubairy}}]{Wang_2008}%
  \BibitemOpen
  \bibfield  {author} {\bibinfo {author} {\bibfnamefont {Li-Gang}\ \bibnamefont
  {Wang}}, \bibinfo {author} {\bibfnamefont {Manzoor}\ \bibnamefont {Ikram}}, \
  and\ \bibinfo {author} {\bibfnamefont {M.~Suhail}\ \bibnamefont {Zubairy}},\
  }\bibfield  {title} {\enquote {\bibinfo {title} {{Control of the
  {Goos-H\"{a}nchen} shift of a light beam via a coherent driving field}},}\
  }\href {\doibase 10.1103/PhysRevA.77.023811} {\bibfield  {journal} {\bibinfo
  {journal} {Phys. Rev. A}\ }\textbf {\bibinfo {volume} {77}},\ \bibinfo
  {pages} {023811} (\bibinfo {year} {2008})}\BibitemShut {NoStop}%
\bibitem [{\citenamefont {Antoni}\ \emph {et~al.}(2011)\citenamefont {Antoni},
  \citenamefont {Kuhn}, \citenamefont {Briant}, \citenamefont {Cohadon},
  \citenamefont {Heidmann}, \citenamefont {Braive}, \citenamefont {Beveratos},
  \citenamefont {Abram}, \citenamefont {Gratiet}, \citenamefont {Sagnes},\ and\
  \citenamefont {Robert-Philip}}]{Antoni2011}%
  \BibitemOpen
  \bibfield  {author} {\bibinfo {author} {\bibfnamefont {Thomas}\ \bibnamefont
  {Antoni}}, \bibinfo {author} {\bibfnamefont {Aur\'{e}lien~G.}\ \bibnamefont
  {Kuhn}}, \bibinfo {author} {\bibfnamefont {Tristan}\ \bibnamefont {Briant}},
  \bibinfo {author} {\bibfnamefont {Pierre-Fran\c{c}ois}\ \bibnamefont
  {Cohadon}}, \bibinfo {author} {\bibfnamefont {Antoine}\ \bibnamefont
  {Heidmann}}, \bibinfo {author} {\bibfnamefont {R\'{e}my}\ \bibnamefont
  {Braive}}, \bibinfo {author} {\bibfnamefont {Alexios}\ \bibnamefont
  {Beveratos}}, \bibinfo {author} {\bibfnamefont {Izo}\ \bibnamefont {Abram}},
  \bibinfo {author} {\bibfnamefont {Luc~Le}\ \bibnamefont {Gratiet}}, \bibinfo
  {author} {\bibfnamefont {Isabelle}\ \bibnamefont {Sagnes}}, \ and\ \bibinfo
  {author} {\bibfnamefont {Isabelle}\ \bibnamefont {Robert-Philip}},\
  }\bibfield  {title} {\enquote {\bibinfo {title} {Deformable two-dimensional
  photonic crystal slab for cavity optomechanics},}\ }\href {\doibase
  10.1364/OL.36.003434} {\bibfield  {journal} {\bibinfo  {journal} {Opt.
  Lett.}\ }\textbf {\bibinfo {volume} {36}},\ \bibinfo {pages} {3434} (\bibinfo
  {year} {2011})}\BibitemShut {NoStop}%
\bibitem [{\citenamefont {Arcizet}\ \emph {et~al.}(2006)\citenamefont
  {Arcizet}, \citenamefont {Cohadon}, \citenamefont {Briant}, \citenamefont
  {Pinard},\ and\ \citenamefont {Heidmann}}]{Arcizet2006}%
  \BibitemOpen
  \bibfield  {author} {\bibinfo {author} {\bibfnamefont {O.}~\bibnamefont
  {Arcizet}}, \bibinfo {author} {\bibfnamefont {P.~F.}\ \bibnamefont
  {Cohadon}}, \bibinfo {author} {\bibfnamefont {T.}~\bibnamefont {Briant}},
  \bibinfo {author} {\bibfnamefont {M.}~\bibnamefont {Pinard}}, \ and\ \bibinfo
  {author} {\bibfnamefont {A.}~\bibnamefont {Heidmann}},\ }\bibfield  {title}
  {\enquote {\bibinfo {title} {Radiation-pressure cooling and optomechanical
  instability of a micromirror},}\ }\href {\doibase 10.1038/nature05244}
  {\bibfield  {journal} {\bibinfo  {journal} {Nature}\ }\textbf {\bibinfo
  {volume} {444}},\ \bibinfo {pages} {71} (\bibinfo {year} {2006})}\BibitemShut
  {NoStop}%
\bibitem [{\citenamefont {Fahad}\ \emph {et~al.}(2026)\citenamefont {Fahad},
  \citenamefont {Shah},\ and\ \citenamefont {Xianlong}}]{fahad2025PSHE}%
  \BibitemOpen
  \bibfield  {author} {\bibinfo {author} {\bibfnamefont {Shah}\ \bibnamefont
  {Fahad}}, \bibinfo {author} {\bibfnamefont {Muzamil}\ \bibnamefont {Shah}}, \
  and\ \bibinfo {author} {\bibfnamefont {Gao}\ \bibnamefont {Xianlong}},\
  }\bibfield  {title} {\enquote {\bibinfo {title} {{Photonic spin Hall effect
  in $\mathcal{PT}$-symmetric non-Hermitian cavity magnomechanics}},}\ }\href
  {\doibase 10.1103/3wks-k8dx} {\bibfield  {journal} {\bibinfo  {journal}
  {Phys. Rev. A}\ }\textbf {\bibinfo {volume} {113}},\ \bibinfo {pages}
  {033524} (\bibinfo {year} {2026})}\BibitemShut {NoStop}%
\end{thebibliography}%
	
\end{document}